\documentclass[sigconf,nonacm]{acmart}

\renewcommand\footnotetextcopyrightpermission[1]{}
\AtBeginDocument{%
  }

\usepackage{amsmath}
\usepackage{multirow}

\usepackage{booktabs}
\usepackage{rotating}
\usepackage{graphicx}

\usepackage{tikz}
\usepackage{pgfplots}
\pgfplotsset{compat=1.18}
\usepackage{subcaption}

\newcommand{\cat}{\mathcal{I}}
\newcommand{\hist}{\mathcal{H}_u}

\newcommand{\epic}{\texttt{EPIC}}

\newcommand{\boldparagraph}[1]{%
  \paragraph{\textnormal{\textbf{#1}}}%
}

\acmConference{}
\acmBooktitle{}
\acmYear{}
\copyrightyear{}
\acmDOI{}
\acmISBN{}

\hypersetup{
  pdfpublisher={The authors},
  pdfpublication={Preprint},
  pdfpubtype={article}
}

\begin{document}

\title[\epic]{\epic: Explicit Posterior Item Conditioning for Semantic ID Diffusion Recommendation}

\author{Tuan-Binh Tran}
\affiliation{%
  \department{College of Engineering and Computer Science}
  \institution{VinUniversity}
  \city{Hanoi}
  \country{Vietnam}
}
\email{binh.tt2@vinuni.edu.vn}

\author{Thanh Tam Nguyen}
\affiliation{%
  \department{School of Information and Communication Technology}
  \institution{Griffith University}
  \city{Gold Coast}
  \state{Queensland}
  \country{Australia}
}
\email{t.nguyen19@griffith.edu.au}

\author{Quoc Viet Hung Nguyen}
\affiliation{%
  \department{School of Information and Communication Technology}
  \institution{Griffith University}
  \city{Gold Coast}
  \state{Queensland}
  \country{Australia}
}
\email{henry.nguyen@griffith.edu.au}

\author{Dung D. Le}
\affiliation{%
  \department{College of Engineering and Computer Science}
  \institution{VinUniversity}
  \city{Hanoi}
  \country{Vietnam}
}
\email{dung.ld@vinuni.edu.vn}

\author{Tung Kieu}
\affiliation{%
  \department{Department of Computer Science}
  \institution{Aalborg University}
  \city{Copenhagen}
  \country{Denmark}
}
\email{tungkvt@cs.aau.dk}

\author{Thanh Trung Huynh}
\authornote{Corresponding author.}
\affiliation{%
  \department{College of Engineering and Computer Science}
  \institution{VinUniversity}
  \city{Hanoi}
  \country{Vietnam}
}
\email{trung.ht@vinuni.edu.vn}
\renewcommand{\shortauthors}{Tran et al.}

\begin{abstract}
Semantic ID (SID) generative recommendation predicts the next item by generating a short tuple of discrete tokens. Recent masked-diffusion methods improve this process through bidirectional context and flexible decoding, yet recommendation ultimately requires selecting among complete catalog items. At each denoising step, a partial SID can correspond to multiple feasible items, while existing methods primarily reason through position-wise token predictions. We propose \textbf{Explicit Posterior Item Conditioning} (\textbf{EPIC}), which introduces explicit item-level competition into SID denoising. \epic{} constructs a personalized posterior over feasible candidate items using the current generation context and the user’s recent interactions, then projects this distribution back to unresolved SID positions to guide subsequent token decisions. The pretrained backbone remains frozen and requires no additional decoder forward pass. Experiments on four Amazon benchmarks show consistent improvements over strong baselines, while diagnostic analyses indicate that the gains primarily arise from personalized transition evidence that preserves promising item hypotheses during denoising.
\end{abstract}

\ccsdesc[500]{Information systems~Recommender systems}
\ccsdesc[300]{Information systems~Retrieval models and ranking}

\keywords{generative recommendation, semantic IDs, masked diffusion,
  sequential recommendation, item-level modeling}

\maketitle

\section{Introduction}

Sequential recommendation predicts the next item a user will interact with from a chronologically ordered history. Classical sequential recommenders score catalog items directly, such that both learning and evaluation are defined at item granularity~\cite{kang2018sasrec, sun2019bert4rec}. Generative recommendation instead formulates the output as a sequence, such as natural-language text or an item identifier~\cite{geng2022recommendation}. A prominent realization of this paradigm represents each item by a Semantic ID (SID), a short tuple of discrete codes obtained by quantizing its content representation, and recommends an item by generating its SID~\cite{rajput2023tiger}. Because related items can share codes, SIDs support parameter sharing across large catalogs while retaining a compact generative interface.

\begin{figure}[t]
\centering
\includegraphics[width=\columnwidth]{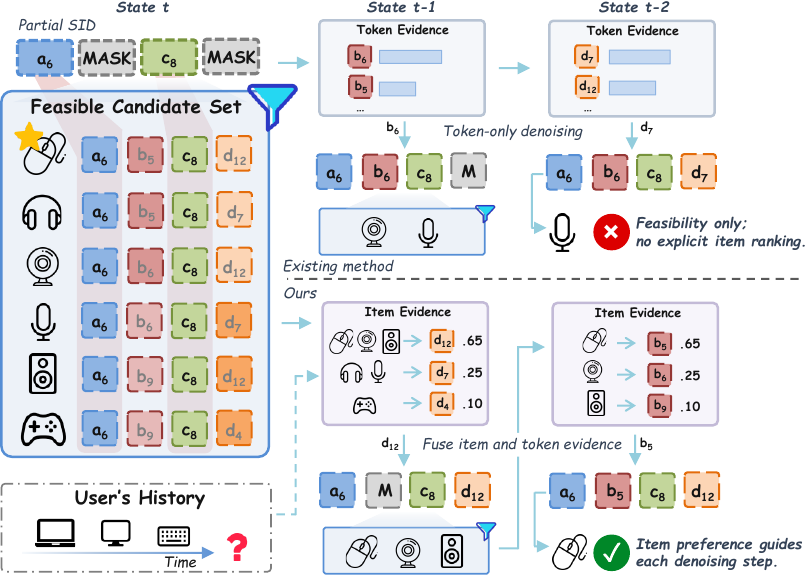}
\caption{\textbf{Why item evidence must act during denoising.} A partial SID defines a feasible candidate set that contracts with each token commitment.
\textit{Top:} token-only denoising follows locally plausible evidence but can eliminate the starred target, making it unreachable for later generation or reranking.
\textit{Bottom:} \epic{} uses the user's recent interactions to compare feasible candidates and maps the resulting item posterior to unresolved SID positions, guiding token commitments while the target remains reachable.
The star denotes the ground-truth item for illustration only.}
\Description{Token-only and \epic{} denoising from a partial SID. Token-only evidence can eliminate the starred target, whereas \epic{} scores feasible candidates using recent interactions and maps the item posterior to unresolved positions to preserve the target.}
\label{fig:motivation}
\end{figure}

Early SID recommenders generate these codes autoregressively, so each decision conditions only on the preceding codes and an early error constrains all subsequent predictions~\cite{rajput2023tiger}. Recent masked-diffusion methods replace this fixed left-to-right process with iterative denoising: they learn to recover masked SID positions from bidirectional context and can resolve positions in a confidence-driven order, potentially several at once~\cite{shi2025lladarec, liu2026diffgrm, shah2025maskgr, mu2026mdgr}. These developments substantially improve how SID tokens are generated.
However, they retain a token-level decision interface even though the recommended object is a complete catalog item.

Figure~\ref{fig:motivation} illustrates the resulting gap. At any intermediate denoising step, the resolved SID positions restrict the catalog to a \emph{feasible candidate set}: the complete items whose identifiers remain consistent with the current state. Feasibility alone does not indicate which candidate best matches the user's preference. Token-level decoding determines how this set contracts through position-wise predictions, leaving competition among the feasible complete items implicit. This matters because every token commitment is simultaneously an item-level elimination decision. A locally plausible code can remove a promising item from the feasible set; once removed, that item is unreachable to subsequent denoising and cannot be recovered by post-hoc reranking over the generated candidates. Item-level evidence is therefore most useful \emph{while} candidates still compete and remain reachable.

Motivated by this observation, we propose \textbf{\underline{E}xplicit \underline{P}osterior \underline{I}tem \underline{C}onditioning (\epic{})}, which introduces explicit item-level competition into Semantic ID denoising. At each step, \epic{} identifies the feasible item hypotheses and constructs a normalized item posterior over a retained candidate support using the current backbone context and the user's recent interactions. It then marginalizes this posterior back to the unresolved SID positions and fuses the resulting item-level evidence with the backbone predictions, allowing preference over complete items to influence subsequent token decisions. As denoising proceeds, the feasible set and posterior are recomputed, so item-level evidence guides which hypotheses remain reachable throughout the generation trajectory.

Two components make this interaction effective. First, \emph{candidate-conditioned transition evidence} compares each candidate with the user's recent complete items, allowing different hypotheses to receive support from different parts of the interaction history. Second, \emph{frontier-aware learning} concentrates direct item-level supervision on states where multiple feasible candidates genuinely compete. The resulting module is lightweight and additive: the pretrained backbone remains frozen, and \epic{} reuses its existing hidden states, token representations, and logits without requiring an additional backbone forward pass.

Experiments on four Amazon benchmarks show that \epic{} consistently improves full-catalog next-item recommendation over strong baselines. Further analyses attribute the gains primarily to personalized transition evidence acting during denoising: it preserves promising candidates as the feasible set contracts, and it depends on the correct user's recent interactions rather than a generic catalog-level signal.

In summary, our main contributions are as follows:

\begin{itemize}
\item We identify the token--item inference gap in masked SID diffusion: every partial SID induces a feasible set of complete-item hypotheses, yet their personalized competition remains implicit in token-level denoising.

\item We propose \epic{}, which constructs an explicit item posterior from candidate-conditioned transition evidence and feeds its item-level marginals back to unresolved SID positions, with frontier-aware learning focusing supervision on states where candidate competition is informative.

\item We demonstrate consistent improvements on four Amazon benchmarks and show that personalized transition evidence is the primary source of the gains, acting by preserving promising candidates throughout denoising.
\end{itemize}

Our full implementation and reproducibility scripts are publicly available.\footnote{\url{https://anonymous.4open.science/r/EPIC/}}
\section{Related Work}

\noindent\textbf{Semantic ID Generative Recommendation.}
Generative retrieval replaces the conventional score-and-index interface with sequence generation, where a model directly emits the identifier of a document or entity and constrained decoding ensures valid outputs~\cite{cao2021genre, tay2022dsi}. \texttt{TIGER} extends this paradigm to recommendation by assigning each item a semantic identifier through residual quantization of content embeddings and autoregressively generating the next item's identifier~\cite{rajput2023tiger}. Subsequent work has explored SID construction from several directions. Vector-quantized text representations~\cite{oord2017vqvae, ni2022sentencet5} support transferable recommendation across domains~\cite{hou2023vqrec}; learnable tokenizers incorporate collaborative signals and reduce code-assignment bias~\cite{wang2024letter, zheng2024lcrec}; contextual tokenization adapts identifiers to the interaction sequence~\cite{hou2025actionpiece}; and longer unordered identifiers trade collision reduction against decoding complexity~\cite{hou2025rpg}. Generative recommendation has also been demonstrated at industrial scale~\cite{zhai2024hstu}. Despite these advances, the common interface remains token-based: preference over a complete item is induced through the generation of its constituent SID tokens rather than represented explicitly during generation.

\vspace{3pt}
\noindent\textbf{Masked and Diffusion Generative Recommendation.}
Discrete diffusion replaces fixed left-to-right decoding with iterative denoising from a masked state~\cite{austin2021d3pm}, an approach recently extended to large language models~\cite{nie2025llada}. A separate line applies continuous diffusion to recommendation by denoising interaction representations or generating target item embeddings~\cite{wang2023diffrec, yang2023dreamrec}; these methods operate outside the SID generation setting considered here. For SIDs, \texttt{LLaDA-Rec} addresses the unidirectional context and error propagation of autoregressive decoding by learning a bidirectional model to recover masked target tokens~\cite{shi2025lladarec}. \texttt{DiffGRM} similarly exploits bidirectional dependencies among codes that jointly identify an item~\cite{liu2026diffgrm}, while \texttt{MaskGR} and \texttt{MDGR} further develop masked-token objectives with parallel or adaptive decoding~\cite{shah2025maskgr, mu2026mdgr}. Related work also improves the history representation itself: masked history learning reconstructs historical items to capture a broader behavioral trajectory~\cite{wei2026mhl}. These methods substantially improve SID generation and condition their token predictors on user interaction histories. However, their training and inference remain primarily parameterized at the token level: supervision is applied to masked positions, and decoding proceeds from per-position vocabulary distributions. Consequently, competition among the complete items compatible with a partial identifier is not explicitly parameterized as a normalized item-level distribution during the denoising process.

\vspace{3pt}
\noindent\textbf{Item-Level Modeling in Generative Recommendation.}
Recent studies have highlighted the limitations of relying solely on token-level generation for item ranking. \texttt{SimGR} shows theoretically and empirically that item-preference distributions induced from generated tokens can be incomplete or distorted, and therefore replaces identifier generation with direct item scoring~\cite{zhao2026simgr}. \texttt{Gryphon} retains the generative interface but applies item-level rescoring after candidate identifiers have been generated~\cite{tikhonovich2026gryphon}. Concurrently, \texttt{ISD} constructs a user-specific item ranking before generation and uses it to support autoregressive SID prefixes prior to beam pruning~\cite{wang2026understanding}. Another related line focuses on output validity, using trie- or graph-constrained decoding to restrict generation to valid catalog identifiers~\cite{cao2021genre, hou2025rpg, shah2025maskgr}. These approaches differ from \epic{} in when and how item-level information is introduced. \texttt{SimGR} bypasses SID generation, \texttt{Gryphon} applies item evidence after candidate generation, and \texttt{ISD} guides autoregressive decoding with a ranking fixed before generation. In contrast, \epic{} operates on arbitrary non-prefix states arising during masked diffusion, recomputes item-level evidence from the current partial SID and user history at each denoising step, and feeds this evidence back to unresolved positions before the next token commitment.

\section{Preliminaries and Problem Formulation}
\label{sec:preliminaries}

\paragraph{Sequential recommendation.}
Let $\cat$ denote the item catalog.  For a user $u$, let $\hist=(i_1,\ldots,i_L)$ be the chronologically ordered interaction history, where $i_\ell\in\cat$.  Sequential recommendation aims to rank the next item $i^\star$ given $\hist$.

\paragraph{Semantic ID generative recommendation.}
Each catalog item $i\in\cat$ is represented by an $H$-tuple of discrete codes,
\begin{equation}
  i \longleftrightarrow
  \mathbf{s}_i=(s_i^1,\ldots,s_i^H),
  \qquad s_i^h\in\mathcal V^h,
  \label{eq:sid}
\end{equation}
obtained by quantizing an item representation~\cite{rajput2023tiger,hou2025rpg}. The mapping from items to tuples need not be injective: two items may collide under the fixed tokenizer. The catalog therefore induces the finite set of distinct complete tuples
\begin{equation}
  \mathcal S_{\cat}
  =\{\mathbf{s}_i:i\in\cat\}
  \subseteq \mathcal V^1\times\cdots\times\mathcal V^H.
\end{equation}
A generative recommender predicts $i^\star$ by generating
$\mathbf{s}_{i^\star}$.  A generated tuple identifies a unique item only when its catalog preimage is a singleton.  Importantly, the valid catalog tuple forms only a small subset of the combinatorial product space.

\subsection{Masked Semantic ID Generation}
\label{sec:prelim-masked-generation}

A masked-diffusion recommender treats the target tuple as a sequence to be iteratively denoised. During training, a sampled subset of target positions is replaced with \texttt{MASK}, and the model learns to recover the original codes. At inference, generation starts from a fully masked target and progressively resolves positions over $T$ denoising steps. At step $t$, the current target state is
\begin{equation}
  \mathbf{x}_t=(x_t^1,\ldots,x_t^H).
\end{equation}
Let
\begin{equation}
  \mathcal O_t=\{h:x_t^h\neq\texttt{MASK}\},
  \qquad
  \mathcal M_t=\{1,\ldots,H\}\setminus\mathcal O_t
\end{equation}
denote its resolved and masked positions. Given $\mathbf{x}_t$ and the user history, the masked backbone produces target hidden states $\mathbf{h}_t^h$ and token logits $\boldsymbol\ell_t^h$. The corresponding token distribution is
\begin{equation}
  p_{\theta,t}^h(v)
  =\operatorname{softmax}(\boldsymbol\ell_t^h)_v,
  \qquad h\in\mathcal M_t.
  \label{eq:backbone-token-posterior}
\end{equation}
The base decoder then uses these distributions to select both the codes and the positions to resolve, producing $\mathbf{x}_{t-1}$.  We use a pretrained masked-diffusion recommender as this token-level backbone and keep it frozen when learning our item-level module.

\subsection{Partial States as Feasible Candidate Sets}
\label{sec:prelim-feasible-set}

Although the backbone predicts individual codes, every partial state already defines a set of complete item hypotheses. Intuitively, once some positions of the target tuple are resolved, only the catalog items whose SIDs agree with those positions can still be the final recommendation.  Formally, we call
\begin{equation}
  \mathcal C_t
  =\left\{
      i\in\cat:
      s_i^h=x_t^h,\ \forall h\in\mathcal O_t
    \right\}
  \label{eq:feasible-candidate-set}
\end{equation}
the \textit{feasible candidate set}. Feasibility here means only SID consistency; it does not imply personalized relevance. Because the catalog and SID table are fixed, $\mathcal C_t$ is obtained exactly by matching resolved positions against all catalog tuples. For masked diffusion, $\mathcal O_t$ may be any subset of positions rather than a prefix.

The feasible candidate set is the item-hypothesis space induced by the current state. When $|\mathcal C_t|>1$, several items remain indistinguishable from the resolved codes; when $|\mathcal C_t|=1$, the item is already identified even if redundant codes remain masked. A fully resolved tuple can still have $|\mathcal C_t|>1$ under a tokenizer collision.

\subsection{The Token--Item Inference Gap}
\label{sec:token-item-gap}

The token distributions in Equation~\eqref{eq:backbone-token-posterior} and
the support in Equation~\eqref{eq:feasible-candidate-set} do not constitute a
distribution over complete items:
\begin{equation}
  p_{\theta,t}^h(v)
  \quad\neq\quad
  P(i\mid\mathbf{x}_t, \hist).
\end{equation}
The first quantity represents uncertainty at one SID position. The second must compare complete candidates, incorporate which candidate is a plausible continuation of this user's behavior, and normalize that competition over the current item-hypothesis space. Further, support restriction only determines which candidates remain feasible; it does not distinguish their personalized preference.  Our goal is therefore to construct an explicit, learned item posterior during denoising and to make that posterior actionable for the token-level generator.

\begin{figure*}[t]
  \centering
  \includegraphics[width=\textwidth]{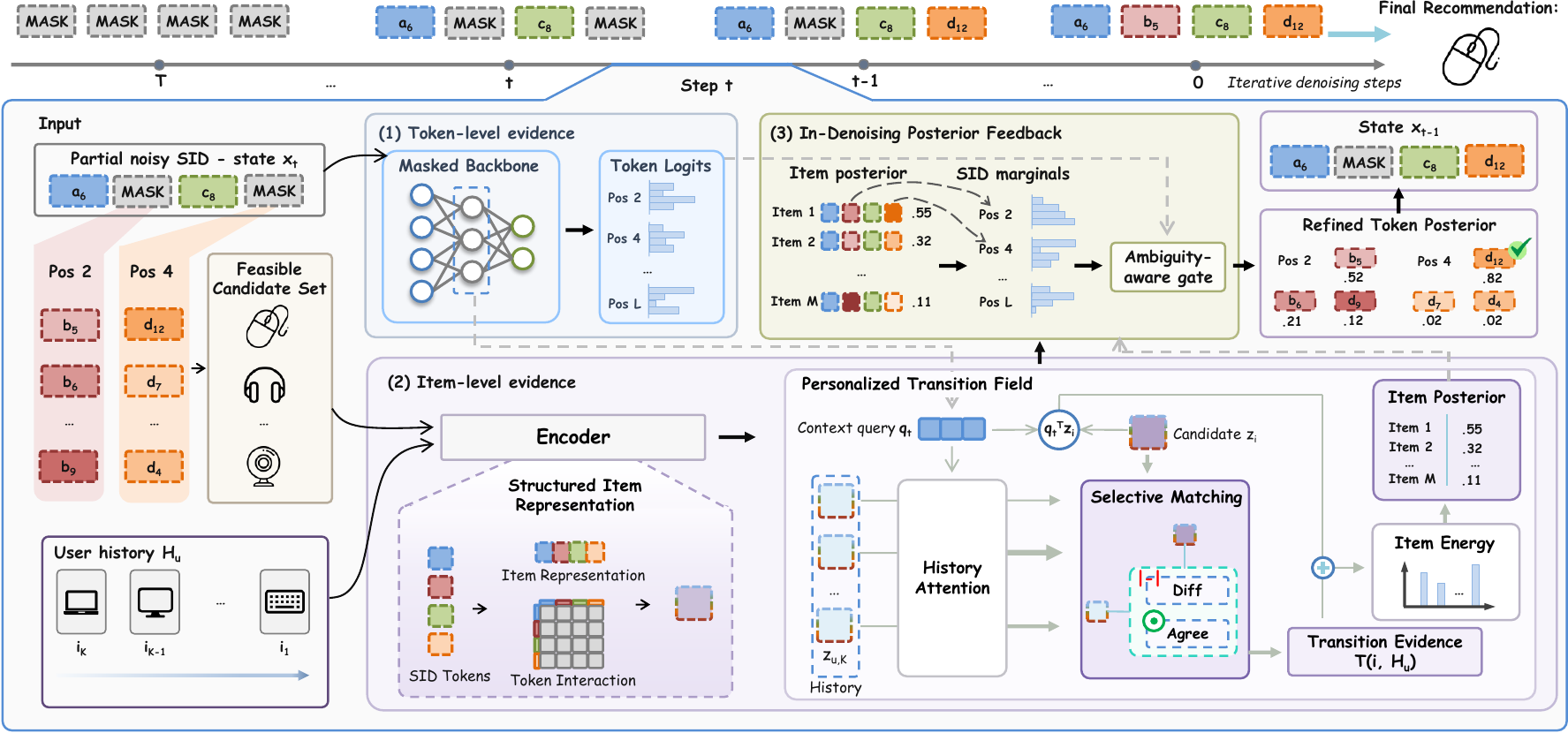}
  \caption{\textbf{\epic{} within iterative SID denoising.} At state $\mathbf x_t$, the frozen backbone provides token evidence, while the item-level branch encodes a retained support of feasible candidates and the user's recent history. Personalized transition evidence yields an item posterior, which is marginalized to unresolved SID positions and fused with the backbone logits through an ambiguity-aware gate. The refined token posterior produces $\mathbf x_{t-1}$, and the procedure repeats.}
  \Description{A partial Semantic ID and user history enter token- and item-level branches. The item branch encodes feasible candidates, selectively matches them with recent items, and produces an item posterior. Its SID marginals are fused with backbone token evidence before the next denoising update.}
  \label{fig:framework}
\end{figure*}

\section{Explicit Posterior Item Conditioning}
\label{sec:method}

\subsection{Framework Overview}
\label{sec:method-overview}

Figure~\ref{fig:framework} follows one denoising step. From the feasible candidate set $\mathcal C_t$ induced by the partial SID, \epic{} retains at most $M$ candidates, constructs complete-item and history representations, and scores each candidate using both the current backbone context and personalized transition evidence. The normalized item posterior is projected to SID codes and injected into the token logits before the ordinary transfer decision, and the updated partial state re-enters the same loop at the next step.

The design follows two principles. First, item-level reasoning is strictly additive: the backbone remains frozen, and every new component, namely the pairwise item encoder, the history path, the transition field, and the feedback gate, enters through a residual with a zero-initialized scale. The initial model therefore recovers backbone inference exactly and departs from it only to the extent that item evidence proves useful during training. Second, item evidence is exact where it matters: the feasible candidate set is computed exactly at every step, and direct item supervision is restricted to states where the posterior is normalized over the complete set rather than a truncation, as detailed in Section~\ref{sec:item-learning}.

\subsection{Item-Level Posterior Construction}
\label{sec:item-level-posterior}

\subsubsection{Feasible candidates and retained support}
\label{sec:candidate-support}

Scoring the full catalog at every step is unnecessary because resolved SID positions already restrict the hypotheses to $\mathcal C_t$. We use the frozen backbone as a non-differentiable proposal mechanism. For $i\in\mathcal C_t$,
\begin{align}
  B_t(i)
  &=\sum_{h\in\mathcal M_t}\log p_{\theta,t}^h(s_i^h),
  \label{eq:token-tuple-score}\\
  \mathcal S_t
  &=\operatorname{TopM}_{i\in\mathcal C_t}B_t(i),
  \qquad |\mathcal S_t|\le M .
  \label{eq:candidate-support}
\end{align}
We call $\mathcal S_t$ the \textit{retained candidate support}; it equals $\mathcal C_t$ whenever $|\mathcal C_t|\le M$. Gradients do not pass through Equation~\eqref{eq:candidate-support}. Thus, token likelihood determines which hypotheses are scored, but not their learned posterior ranking. This separation is deliberate: it prevents the same token evidence from being counted twice, once in proposal formation and again in the posterior energy, and Section~\ref{sec:ablation} shows that folding token likelihood into the posterior is consistently harmful. Intuitively, token likelihood is a reliable instrument for pruning hypotheses but a poor one for ranking the survivors. The feasibility-only rung in the capability ladder uses the same $\mathcal C_t$ restriction but omits the learned item posterior and transition field.

\subsubsection{Structured item and history representations}
\label{sec:structured-item-representation}

Candidates and historical items share an encoder over complete SIDs, so both sides of the later comparison live in one representation space. Reusing the frozen code embeddings also introduces no new item vocabulary: an item is represented purely through the codes the backbone already understands. Let $\mathbf c_i^h=\mathbf w(s_i^h)+\mathbf d_h$ combine the frozen code embedding with a learned position embedding, and let $\bar{\mathbf z}_i=f_{\mathrm{tuple}} ([\mathbf c_i^1;\ldots;\mathbf c_i^H])$. Concatenation alone, however, does not represent interactions among SID heads, which describe complementary facets of the same item. With $\mathbf u_i^h=W_p\mathbf c_i^h$ we therefore add
\begin{align}
  \mathbf m_i
  &=
  \frac{\left(\sum_h\mathbf u_i^h\right)^{\odot2}
  -\sum_h(\mathbf u_i^h)^{\odot2}}{H(H-1)},\label{eq:pairwise-summary}\\
  \mathbf z_i
  &=\operatorname{norm}\!\left(
  \bar{\mathbf z}_i+\tanh(\gamma_p)f_p(\mathbf m_i)\right).
  \label{eq:structured-item-encoder}
\end{align}
This computes the mean pairwise SID-head interaction in $O(H)$ time.
The same encoder maps the $K$ most recent items to $\mathbf z_{u,1},\ldots,\mathbf z_{u,K}$, ordered from newest to oldest. Items sharing a complete SID remain tied because \epic{} adds no item-identity feature.

The backbone states supply the current denoising context, while a lightweight GRU summarizes the retained history window:
\begin{equation}
  \mathbf q_t=\operatorname{norm}\!\left[
  f_q\!\left(\frac{1}{H}\sum_{h=1}^{H}\mathbf h_t^h\right)
  +\tanh(\gamma_h)
  \operatorname{GRU}(\mathbf z_{u,K},\ldots,\mathbf z_{u,1})
  \right].
  \label{eq:item-query}
\end{equation}
The GRU consumes items chronologically from oldest to newest; $\mathbf q_t$ is shared when scoring all candidates in $\mathcal S_t$.

\subsubsection{Personalized transition field and item posterior}
\label{sec:transition-field}
\label{sec:explicit-item-posterior}

As illustrated by the selective-matching block in Figure~\ref{fig:framework}, history attention determines which recent interactions matter at the current state, while the candidate--history interaction determines how each candidate relates to them:
\begin{align}
  a_{t,r}
  &=\operatorname{softmax}_{r}
    \left(\mathbf q_t^\top\mathbf z_{u,r}/\tau_H+b_r\right),
  \label{eq:history-attention}\\
  \phi(i,r)
  &=f_{\mathrm{trans}}\!\left(
    [\,\mathbf z_i\odot\mathbf z_{u,r};
       |\mathbf z_i-\mathbf z_{u,r}|\,]\right),
  \label{eq:candidate-history-interaction}\\
  T_t(i,\hist)
  &=\sum_{r=1}^{K}a_{t,r}\phi(i,r).
  \label{eq:transition-evidence}
\end{align}
Here $b_r$ is a learned recency bias for the $r$-th newest item, so the shape of the recency prior is estimated from data rather than imposed as a fixed temporal decay. Although $a_{t,r}$ is shared across candidates, the agreement/difference features in $\phi(i,r)$ make $T_t(i,\hist)$ candidate-specific: different hypotheses can draw support from different parts of the recent history, which a single pooled history representation cannot express.

We combine direct context relevance with transition evidence and normalize over complete candidates:
\begin{align}
  E_t(i)
  &=\frac{\mathbf q_t^\top\mathbf z_i
    +\tanh(\gamma_T)T_t(i,\hist)}{\kappa},
  \label{eq:item-energy}\\
  \pi_t(i)
  &=\frac{\mathbf 1[i\in\mathcal S_t]\exp E_t(i)}
    {\sum_{j\in\mathcal S_t}\exp E_t(j)}.
  \label{eq:item-posterior}
\end{align}
Thus $\pi_t$ explicitly represents competition among complete items in the retained support and covers all feasible candidates when $|\mathcal C_t|\le M$. Here, posterior denotes the normalized learned distribution in Equation~\ref{eq:item-posterior}. Importantly, $B_t(i)$ forms $\mathcal S_t$ but is not added to $E_t(i)$; the principal posterior is learned independently of token likelihood. Relaxing this separation yields the product-of-experts and cavity alternatives evaluated in the ablation study, and setting $\gamma_T=0$ gives the generic-posterior rung, while the full model includes transition memory.

\subsubsection{Frontier-aware learning}
\label{sec:item-learning}

Direct item supervision is most informative when several exact hypotheses compete. We therefore apply it only on the identification frontier $w_t=\mathbf 1[\,2\le|\mathcal C_t|\le M\,]$. The lower bound removes singleton states, where the item is already determined and there is nothing to discriminate; the upper bound guarantees $\mathcal S_t=\mathcal C_t$, so the supervised posterior is normalized over every feasible candidate rather than a truncated proposal set. At heavily masked states, however, insisting on the exact target is a noisy objective, since many candidates share most of the target's SID structure. To retain this structure under heavy masking, let
\begin{align}
  d_{\mathrm{SID}}(i,i^\star)
  &=\frac{1}{H}\sum_{h=1}^{H}
    \mathbf 1[s_i^h\ne s_{i^\star}^h],\\
  r_t(i\mid i^\star)
  &=\frac{\exp[-d_{\mathrm{SID}}(i,i^\star)/\tau_S]}
    {\sum_{j\in\mathcal C_t}
     \exp[-d_{\mathrm{SID}}(j,i^\star)/\tau_S]} .
  \label{eq:sid-kernel-target}
\end{align}
With masked fraction $\rho_t=|\mathcal M_t|/H$, the objectives are
\begin{align}
  \mathcal L_{\mathrm{item}}
  &=w_t\!\left[-\log\pi_t(i^\star)
    +\eta\rho_t\operatorname{CE}(r_t,\pi_t)\right],
  \label{eq:item-loss}\\
  \mathcal L
  &=\mathcal L_{\mathrm{tok}}(\widetilde{\boldsymbol\ell})
    +\beta\mathcal L_{\mathrm{item}} .
  \label{eq:total-objective}
\end{align}
The $\rho_t$ scaling makes the structured smoothing strongest when the target is most corrupted and lets it vanish as the SID resolves, so the objective anneals toward exact-item discrimination. The pairwise residual, transition term, frontier restriction, and SID kernel are separable components used in the removal study. Only the item-level adapter is updated; the masked-diffusion backbone receives no gradients.

\subsection{In-Denoising Posterior Feedback}
\label{sec:item-to-token-feedback}

\subsubsection{Item-to-token marginalization}
\label{sec:posterior-marginalization}

For unresolved position $h$ and code $v\in\mathcal V^h$, we project the same complete-item posterior to the SID vocabulary:
\begin{equation}
  p_{I,t}^h(v)
  =\sum_{i\in\mathcal S_t}
    \pi_t(i)\mathbf 1[s_i^h=v].
  \label{eq:item-to-token-marginal}
\end{equation}
This is the exact positional marginal of $\pi_t$. It preserves joint item-level competition across positions and assigns mass only to codes used by at least one retained complete item.

\subsubsection{Ambiguity-aware residual fusion}
\label{sec:posterior-feedback}

A straightforward fusion would be a convex mixture of $p_{I,t}^h$ and $p_{\theta,t}^h$. Yet, such a mixture constrains the feedback to be positive and combines two distributions whose calibration may differ. We instead express item guidance as a clipped log-probability residual,
\begin{equation}
  \Delta_t^h(v)=\operatorname{clip}\!\left(
  \log[p_{I,t}^h(v)+\epsilon]
  -\log[p_{\theta,t}^h(v)+\epsilon],-c,c\right),
  \label{eq:posterior-residual}
\end{equation}
where $\epsilon>0$ handles codes absent from the retained support. The gate uses the target state $\mathbf h_t^h$, normalized feasible-set size $A_t=\log\max(1,|\mathcal C_t|)/\log|\cat|$, normalized item-posterior entropy $\bar H(\pi_t)$, top-two posterior margin $m_t$, masked fraction $\rho_t$, and normalized backbone-token entropy $\bar H(p_{\theta,t}^h)$:
\begin{align}
  \mathbf z_t^h
  &=
  [\,\mathbf h_t^h; A_t; \bar H(\pi_t); m_t;
     \rho_t; \bar H(p_{\theta,t}^h)\,],
  \notag\\[-2pt]
  g_t^h
  &=
  \lambda_F \tanh f_g(\mathbf z_t^h)\,
  \mathbf 1[h\in\mathcal M_t],
  \label{eq:ambiguity-gate}\\
  \widetilde\ell_t^h(v)
  &=
  \ell_t^h(v)+g_t^h\Delta_t^h(v).
  \label{eq:refined-token-logits}
\end{align}

We normalize the item- and token-level entropies as
\begin{align*}
  \bar H(\pi_t)
  =
  \frac{H(\pi_t)}
       {\log\!\max\{2,|\mathcal S_t|\}},\\
  \bar H(p_{\theta,t}^h)
  =
  \frac{H(p_{\theta,t}^h)}
       {\log |\mathcal V^h|}.
\end{align*}

Moreover, $m_t=\pi_t^{(1)}-\pi_t^{(2)}$ when $|\mathcal S_t|\ge2$, and $m_t=1$ otherwise. The gate is position-specific and conditioned on both item- and token-level uncertainty because the reliability of item evidence changes throughout denoising: early states carry broad feasible sets and diffuse posteriors, whereas late states may already be determined by the resolved codes. The signed gate can promote or suppress item evidence accordingly, and its zero-initialized final layer makes feedback an exact initial no-op.

\boldparagraph{Iterative inference and cost.}
The base transfer rule consumes $\operatorname{softmax}(\widetilde{\boldsymbol\ell}_t^h)$ to produce $\mathbf x_{t-1}$, and applying the loop at every step yields a sequence of posteriors $\pi_T,\ldots,\pi_0$ over progressively contracting feasible sets. At each step, $\mathcal C_t$, $\mathcal S_t$, and $\pi_t$ are recomputed from the updated state; no posterior state is carried across denoising steps. Item evidence therefore changes the denoising trajectory itself, which separates \epic{} from both post-hoc reranking and a recurrent latent-posterior model. Regarding cost, the exact feasible-candidate scan costs $O(|\cat|H)$, the candidate--history interaction costs $O(MKd)$, and marginalization costs $O(MH)$ per step. All computations reuse the current backbone states and logits, requiring no additional backbone forward pass; an inverted SID index could replace the dense catalog scan at larger scale.

\section{Experimental Setup}
\label{sec:experimental-setup}

\subsection{Datasets and Evaluation Protocol}

We evaluate full-catalog next-item recommendation on four 5-core Amazon Review categories, which are used by prior SID recommenders~\cite{rajput2023tiger,hou2026bridging}. For each user, interactions are ordered chronologically. The final interaction is held out for testing, the penultimate interaction is used for validation, and all preceding interactions form the training sequence. Table~\ref{tab:datasets} summarizes the resulting datasets.

\begin{table}[t]
  \caption{Dataset statistics after preprocessing.}
  \label{tab:datasets}
  \centering
  \resizebox{\columnwidth}{!}{%
  \begin{tabular}{lrrrr}
    \toprule
    & \textbf{Beauty} & \textbf{Sports} & \textbf{Toys} & \textbf{Musical} \\
    \midrule
    Users                & 22{,}363 & 35{,}598 & 19{,}412 & 57{,}439 \\
    Items                & 12{,}101 & 18{,}357 & 11{,}924 & 24{,}587 \\
    Interactions         & 198{,}502 & 296{,}337 & 167{,}597 & 511{,}836 \\
    Mean sequence length & 8.88 & 8.32 & 8.63 & 8.91 \\
    \bottomrule
  \end{tabular}
  }
\end{table}

\subsection{Baselines and Metrics}

We compare \epic{} with 3 groups of baselines. 
The first group includes Item ID-based sequential recommenders, represented by 
\texttt{BERT4Rec}~\cite{sun2019bert4rec}, 
\texttt{S$^3$-Rec}~\cite{zhou2020s3rec} and
\texttt{HSTU}~\cite{zhai2024hstu}. 
The second group contains SID-based generative baselines, such as 
\texttt{TIGER}~\cite{rajput2023tiger},
\texttt{Action-\\Piece}~\cite{hou2025actionpiece} and
\texttt{RPG}~\cite{hou2025rpg}. 
The third group consists of masked-diffusion generative recommenders, including 
\texttt{MaskGR}~\cite{shah2025maskgr}, 
\texttt{LLaDA}-Rec~\cite{shi2025lladarec} and 
\texttt{DiffGRM}~\cite{liu2026diffgrm}. 
Full baseline citations are given in Table~\ref{tab:overall-four-datasets}.  A $\dagger$ indicates a method for which at least one released implementation was executed locally. We retain a local result only when its data split, full-catalog ranking protocol, seen-item handling, and metric implementation match ours. All remaining baseline values are taken from the cited sources and marked with $\ddagger$.

We report full-catalog Recall@$K$ and NDCG@$K$ for $K\in\{5,10\}$ under a unique-SID item-level protocol. For tokenizers that retain collisions and do not introduce item-specific disambiguation tokens, a predicted SID receives item-level credit only when it is catalog-valid, has a singleton catalog preimage, and maps to the target item. Ambiguous SIDs receive no credit.

Results for \epic{} are reported as mean$\pm$sample standard deviation over five independently trained runs, with checkpoints selected using validation NDCG@5 only. For each of the 16 dataset--metric combinations, we compare the per-user scores of \epic{} with those of the corresponding second-best method using a two-sided paired $t$-test. All second-best methods were reproduced locally under the same evaluation protocol, enabling matched per-user comparisons. We apply Holm correction jointly across the 16 tests; all adjusted comparisons satisfy $p_{\mathrm{adj}}<0.05$.

\subsection{Training and Inference Details}

Following \texttt{LLaDA-Rec}~\cite{shi2025lladarec}, we use its pretrained \texttt{Sentence-T5}-based tokenizer~\cite{ni2022sentencet5}, which maps each item to a four-position SID with 256 codes per position. The tokenizer and the resulting item-to-SID mapping remain fixed and are shared across all matched comparisons.

\epic{} adopts the bidirectional Transformer and four-step transfer schedule of \texttt{LLaDA-Rec}. We freeze the backbone and optimize only the item-reasoning adapter. During inference, \epic{} injects item-level evidence into each denoising step using the procedure described in Section~\ref{sec:method}. All matched interventions use the same tokenizer, backbone decoder, beam width, candidate budget, and completion rule.

We select the backbone depth, learning rate, candidate budget $M$, and recency window $K$ using validation NDCG@5, searching over $L\in\{4,6\}$, $lr\in\{10^{-3},3{\times}10^{-3}\}$, $M\in\{64,128,256\}$, and $K\in\{2,4,8\}$. All remaining settings are shared across datasets: hidden dimension 256, four attention heads, four SID positions and transfer steps, \texttt{AdamW} with weight decay $0.01$ and warmup ratio $0.05$, batch size 512, at most 8{,}000 updates, posterior temperature $\kappa=0.1$, and transition temperature $\tau=0.25$. Training uses early stopping according to validation NDCG@5. Efficiency measurements are conducted on one Tesla T4 using the same software stack and batch configuration for both methods.

\section{Results and Analysis}
We organize our evaluation around four research questions that separate overall effectiveness from the mechanism, source, and cost of EPIC's gains:

\begin{itemize}
\setlength{\itemsep}{1pt}
\setlength{\topsep}{2pt}
    \item[\textbf{RQ1.}] \textit{Overall effectiveness.} Does EPIC improve full-catalog next-item recommendation over conventional, Semantic ID generative, and masked-diffusion baselines across datasets?
    \item[\textbf{RQ2.}] \textit{In-denoising mechanism.} Does item evidence improve recommendation by preserving promising target hypotheses during denoising, rather than merely reranking completed outputs?
    \item[\textbf{RQ3.}] \textit{Source of improvement.} Which model components and personalized history signals account for the gains, and how does the chosen item-posterior construction affect performance?
    \item[\textbf{RQ4.}] \textit{Operating regimes and cost.} Under what ambiguity and partial-resolution regimes is EPIC effective, and what parameter, training, and inference costs does its item-level reasoning introduce?
\end{itemize}
We address RQ1 in Section~\ref{sec:overall}, RQ2 in Sections~\ref{sec:hypotheses} and \ref{sec:case_study}, RQ3 in Section~\ref{sec:ablation}, and RQ4 in Section~\ref{sec:effective}.

\subsection{Overall Performance} \label{sec:overall}
\providecommand{\best}[1]{\textbf{#1}}
\providecommand{\second}[1]{\underline{#1}}
\newcommand{\beststat}[2]{%
  \shortstack{%
    \textbf{#1}\\[-1.2pt]
    \scriptsize$\pm$#2%
  }%
}

\newcommand{\secondstat}[2]{%
  \shortstack{%
    \underline{#1}\\[-1.2pt]
    \scriptsize$\pm$#2%
  }%
}

\begin{table*}[t]
\centering
\caption{Full-catalog next-item recommendation. R@K and N@K denote Recall@K and NDCG@K; best and second-best results are bold and underlined.}
\label{tab:overall-four-datasets}

\vspace{2pt}
\resizebox{\textwidth}{!}{
\begin{tabular}{@{}lcccccccccccccccc@{}}
\toprule
& \multicolumn{4}{c}{\textbf{Sports and Outdoors}}
& \multicolumn{4}{c}{\textbf{Beauty}}
& \multicolumn{4}{c}{\textbf{Toys and Games}}
& \multicolumn{4}{c}{\textbf{Musical Instruments}} \\
\cmidrule(lr){2-5}
\cmidrule(lr){6-9}
\cmidrule(lr){10-13}
\cmidrule(lr){14-17}

\textbf{Model}
& R@5 & N@5 & R@10 & N@10
& R@5 & N@5 & R@10 & N@10
& R@5 & N@5 & R@10 & N@10
& R@5 & N@5 & R@10 & N@10 \\
\midrule

\multicolumn{17}{c}{\textit{Item ID-based Methods}} \\
\midrule

Caser~\cite{tang2018caser}$^\ddagger$
& .0116 & .0072 & .0194 & .0097
& .0205 & .0131 & .0347 & .0176
& .0166 & .0107 & .0270 & .0141
& .0241 & .0151 & .0386 & .0197 \\

GRU4Rec~\cite{hidasi2016gru4rec}$^\ddagger$
& .0129 & .0086 & .0204 & .0110
& .0164 & .0099 & .0283 & .0137
& .0097 & .0059 & .0176 & .0084
& .0324 & .0209 & .0501 & .0266 \\

HGN~\cite{ma2019hgn}$^\ddagger$
& .0189 & .0120 & .0313 & .0159
& .0325 & .0206 & .0512 & .0266
& .0321 & .0221 & .0497 & .0277
& .0321 & .0202 & .0517 & .0265 \\

BERT4Rec~\cite{sun2019bert4rec}$^\ddagger$
& .0115 & .0075 & .0191 & .0099
& .0203 & .0124 & .0347 & .0170
& .0116 & .0071 & .0203 & .0099
& .0307 & .0195 & .0485 & .0252 \\

SASRec~\cite{kang2018sasrec}$^\ddagger$
& .0233 & .0154 & .0350 & .0192
& .0387 & .0249 & .0605 & .0318
& .0463 & .0306 & .0675 & .0374
& .0333 & .0213 & .0523 & .0274 \\

HSTU~\cite{zhai2024hstu}$^\dagger$
& .0274 & .0176 & .0438 & .0229
& \second{.0563} & .0348
& \second{.0853} & .0442
& \second{.0630} & .0390
& \second{.0916} & \second{.0483}
& .0343 & .0221 & .0553 & .0289 \\


FDSA~\cite{zhang2019fdsa}$^\ddagger$
& .0182 & .0122 & .0288 & .0156
& .0267 & .0163 & .0407 & .0208
& .0228 & .0140 & .0381 & .0189
& .0347 & .0230 & .0545 & .0293 \\

S$^3$-Rec~\cite{zhou2020s3rec}$^\ddagger$
& .0251 & .0161 & .0385 & .0204
& .0387 & .0244 & .0647 & .0327
& .0443 & .0294 & .0700 & .0376
& .0317 & .0199 & .0496 & .0257 \\


VQ-Rec~\cite{hou2023vqrec}$^\ddagger$
& .0208 & .0144 & .0300 & .0173
& .0457 & .0317 & .0664 & .0383
& .0497 & .0346 & .0737 & .0423
& .0361 & .0231 & .0578 & .0301 \\

\midrule
\multicolumn{17}{c}{\textit{Semantic ID-based Generative Methods}} \\
\midrule

TIGER~\cite{rajput2023tiger}$^\dagger$
& .0233 & .0149 & .0364 & .0191
& .0382 & .0249 & .0615 & .0323
& .0328 & .0207 & .0514 & .0266
& .0341 & .0221 & .0537 & .0284 \\

LETTER~\cite{wang2024letter}$^\ddagger$
& .0288 & .0198 & .0435 & .0244
& .0500 & .0340 & .0708 & .0406
& .0547 & .0389 & .0741 & .0452
& .0372 & .0246 & .0580 & .0313 \\

LC-Rec~\cite{zheng2024lcrec}$^\ddagger$
& .0276 & .0188 & .0417 & .0233
& .0478 & .0334 & .0675 & .0397
& .0540 & .0384 & .0734 & .0447
& .0379 & .0251 & .0587 & .0318 \\

ActionPiece~\cite{hou2025actionpiece}$^\dagger$
& \second{.0330} & \second{.0224}
& \second{.0500} & \second{.0264}
& .0553 & \second{.0379}
& .0775 & .0424
& .0559 & \second{.0398}
& .0760 & .0463
& .0383 & .0243 & .0615 & .0318 \\

RPG~\cite{hou2025rpg}$^\dagger$
& .0304 & .0211 & .0441 & .0255
& .0533 & .0368 & .0788
& \second{.0450}
& .0574 & .0390 & .0859 & .0481
& .0191 & .0132 & .0272 & .0158 \\

\midrule
\multicolumn{17}{c}{\textit{Diffusion-based Generative Methods}} \\
\midrule


MaskGR~\cite{shah2025maskgr}$^\ddagger$
& .0302 & .0191 & .0454 & .0249
& .0538 & .0351 & .0815 & .0441
& .0548 & .0375 & .0846 & .0445
& -- & -- & -- & -- \\

LLaDA-Rec~\cite{shi2025lladarec}$^\dagger$
& .0287 & .0186 & .0441 & .0235
& .0525 & .0359 & .0770 & .0438
& .0484 & .0328 & .0718 & .0403
& .0405
& \second{.0265}
& \second{.0630}
& \second{.0337} \\

DiffGRM~\cite{liu2026diffgrm}$^\dagger$
& .0183 & .0117 & .0287 & .0150
& .0321 & .0207 & .0498 & .0264
& .0301 & .0208 & .0435 & .0251
& .0237 & .0152 & .0348 & .0189 \\

\midrule

\textbf{EPIC} (Ours)
& \beststat{.0353$^{*}$}{.0003}
& \beststat{.0244$^{*}$}{.0001}
& \beststat{.0521$^{*}$}{.0006}
& \beststat{.0298$^{*}$}{.0002}

& \beststat{.0610$^{*}$}{.0004}
& \beststat{.0426$^{*}$}{.0003}
& \beststat{.0866$^{*}$}{.0002}
& \beststat{.0509$^{*}$}{.0002}

& \beststat{.0662$^{*}$}{.0009}
& \beststat{.0465$^{*}$}{.0007}
& \beststat{.0927$^{*}$}{.0011}
& \beststat{.0551$^{*}$}{.0007}

& \beststat{.0424$^{*}$}{.0001}
& \beststat{.0282$^{*}$}{.0001}
& \beststat{.0640$^{*}$}{.0003}
& \beststat{.0352$^{*}$}{.0001} \\

\addlinespace[2pt]
\textit{Gain vs. second-best}
& +7.0\% & +8.9\% & +4.2\% & +12.9\%
& +8.3\% & +12.4\% & +1.5\% & +13.1\%
& +5.1\% & +16.8\% & +1.2\% & +14.1\%
& +4.7\% & +6.4\% & +1.6\% & +4.5\% \\

\bottomrule
\end{tabular}
}

\vspace{2pt}
\begin{minipage}{\textwidth}
\footnotesize
$\dagger$ denotes locally reproduced results, whereas $\ddagger$ denotes paper-reported results. Citations beside model names identify the original methods. The paper-reported cells were consolidated from the benchmark tables of MHL~\cite{wei2026mhl}, UTGRec~\cite{zheng2025utgrec}, Pctx~\cite{zhong2025pctx}, and LLaDA-Rec~\cite{shi2025lladarec}. Section~\ref{sec:experimental-setup} details the evaluation protocol. A dash denotes an unavailable result. \epic{} reports mean$\pm$sample standard deviation over five independent runs. The gain row gives the relative improvement of the \epic{} mean over the second-best result in each column. $^{*}$ marks a significant improvement over the corresponding second-best method under a two-sided paired $t$-test over users, with Holm correction jointly applied across all 16 dataset--metric comparisons ($p_{\mathrm{adj}}<0.05$).
\end{minipage}

\end{table*}

Table~\ref{tab:overall-four-datasets} summarizes full-catalog next-item recommendation performance on the four benchmarks. \epic{} achieves the best mean in all 16 metric--dataset combinations. Relative to the second-best result in each column, its gains range from 1.2\% to 16.8\%. The largest improvements occur on ranking-sensitive metrics: 16.8\% in NDCG@5 and 14.1\% in NDCG@10 on \textbf{Toys} and \textbf{Games}, 13.1\% in NDCG@10 on \textbf{Beauty}, and 12.9\% in NDCG@10 on \textbf{Sports} and \textbf{Outdoors}. Even against strong locally reproduced competitors, \epic{} remains consistently better.

The improvements are stable across independent training runs, with sample standard deviations of at most .0011 across all reported metrics. Moreover, every gain over the corresponding second-best method remains statistically significant under a two-sided paired $t$-test after Holm correction across all 16 dataset--metric comparisons ($p_{\mathrm{adj}}<0.05$). Together, these results indicate that explicit posterior item conditioning consistently improves both top-ranked accuracy and deeper-list recall.

\subsection{Does Item Evidence Preserve Hypotheses During Denoising?} \label{sec:hypotheses}
\noindent\boldparagraph{Inline versus post-hoc evidence.}
\label{sec:abl-inline-posthoc}

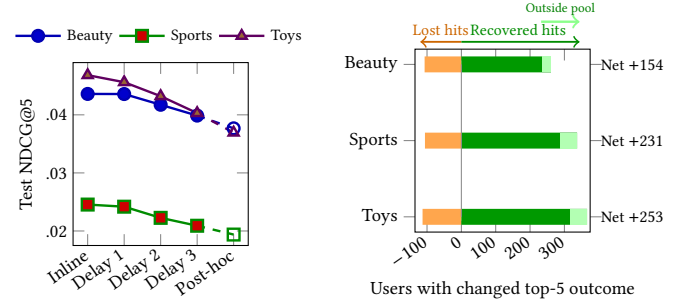
\begin{figure}[t]
\centering
\begin{subfigure}[t]{0.46\linewidth}
  \scriptsize
  \centering
  \begin{tikzpicture}
    \begin{axis}[
      width=\linewidth,
      height=0.55*\axisdefaultheight,
      ylabel={Test NDCG@5},
      symbolic x coords={Inline,S1,S2,S3,Post},
      xtick={Inline,S1,S2,S3,Post},
      xticklabels={Inline,Delay 1,Delay 2,Delay 3,Post-hoc},
      xticklabel style={font=\footnotesize,rotate=35,anchor=east,yshift=-2pt},
      tick label style={font=\footnotesize},
      label style={font=\footnotesize},
      ymin=.0175,ymax=.049,
      scaled y ticks=false,
      ytick={.02,.03,.04},
      yticklabels={.02,.03,.04},
      legend style={at={(0.5,1.05)},anchor=south,legend columns=3,
                    font=\scriptsize,draw=none,fill=none},
      every axis plot/.append style={line width=.9pt,mark size=2.2pt},
    ]
      enlarge x limits=.08,
      \addplot+[blue!70!black,mark=*] coordinates {
        (Inline,.043602)(S1,.043584)(S2,.041735)(S3,.039871)};
      \addlegendentry{Beauty}
      \addplot+[blue!70!black,dashed,no marks,forget plot]
        coordinates {(S3,.039871)(Post,.037676)};
      \addplot+[blue!70!black,only marks,mark=o,mark options={fill=white},forget plot]
        coordinates {(Post,.037676)};
      \addplot+[green!55!black,mark=square*] coordinates {
        (Inline,.024548)(S1,.024183)(S2,.022262)(S3,.020901)};
      \addlegendentry{Sports}
      \addplot+[green!55!black,dashed,no marks,forget plot]
        coordinates {(S3,.020901)(Post,.019384)};
      \addplot+[green!55!black,only marks,mark=square,mark options={fill=white},forget plot]
        coordinates {(Post,.019384)};
      \addplot+[violet!75!black,mark=triangle*] coordinates {
        (Inline,.046834)(S1,.045624)(S2,.043216)(S3,.040303)};
      \addlegendentry{Toys}
      \addplot+[violet!75!black,dashed,no marks,forget plot]
        coordinates {(S3,.040303)(Post,.036936)};
      \addplot+[violet!75!black,only marks,mark=triangle,mark options={fill=white},forget plot]
        coordinates {(Post,.036936)};
    \end{axis}
  \end{tikzpicture}
  \caption{Later feedback lowers NDCG@5.}
  \label{fig:epic-feedback-timing}
\end{subfigure}%
\hspace{0.06\linewidth}%
\begin{subfigure}[t]{0.46\linewidth}
  \scriptsize
  \centering
  \begin{tikzpicture}
    \begin{axis}[
      xbar,
      width=\linewidth,
      height=0.55*\axisdefaultheight,
      xmin=-135,xmax=375,
      xtick={-100,0,100,200,300},
      xticklabel style={font=\footnotesize,rotate=35,anchor=east,yshift=-2pt},
      xlabel={Users with changed top-5 outcome},
      symbolic y coords={Beauty,Sports,Toys},
      ytick=data,
      y dir=reverse,
      tick label style={font=\footnotesize},
      label style={font=\footnotesize},
      bar width=6pt,
      clip=false,
    ]
      \addplot+[fill=orange!70,draw=none,bar shift=0pt]
        coordinates {(-107,Beauty)(-106,Sports)(-113,Toys)};
      \addplot+[fill=green!60!black,draw=none,bar shift=0pt]
        coordinates {(261,Beauty)(337,Sports)(366,Toys)};
      \fill[green!30] ([yshift=-3pt]axis cs:236,Beauty)
        rectangle ([yshift=3pt]axis cs:261,Beauty);
      \fill[green!30] ([yshift=-3pt]axis cs:289,Sports)
        rectangle ([yshift=3pt]axis cs:337,Sports);
      \fill[green!30] ([yshift=-3pt]axis cs:317,Toys)
        rectangle ([yshift=3pt]axis cs:366,Toys);
      \draw[black!45] (axis cs:0,Beauty) -- (axis cs:0,Toys);
      \draw[->,orange!80!black,line width=.7pt]
        (rel axis cs:.265,1.04) -- node[above,font=\scriptsize]{Lost hits}
        (rel axis cs:.03,1.04);
      \draw[->,green!55!black,line width=.7pt]
        (rel axis cs:.265,1.04) -- node[pos=.48,above,font=\scriptsize]{Recovered hits}
        (rel axis cs:.94,1.04);
      \draw[->,green!45,line width=.9pt]
        (rel axis cs:.72,1.15) --
        node[midway,above,font=\tiny,text=green!50!black]{Outside pool}
        (rel axis cs:.94,1.15);
      \node[font=\scriptsize,anchor=west] at (axis cs:390,Beauty) {Net $+154$};
      \node[font=\scriptsize,anchor=west] at (axis cs:390,Sports) {Net $+231$};
      \node[font=\scriptsize,anchor=west] at (axis cs:390,Toys) {Net $+253$};
    \end{axis}
  \end{tikzpicture}
  \caption{Inline feedback recovers more top-5 hits than it loses.}
  \label{fig:epic-hard-rescue}
\end{subfigure}
\caption{\textbf{Inline item evidence changes denoising.} Delayed feedback lowers NDCG@5. Inline feedback recovers more top-5 targets than it loses; light segments identify recoveries absent from the final control pool.}
\label{fig:epic-inline-reasoning}
\vspace{-8pt}
\end{figure}

Figure~\ref{fig:epic-inline-reasoning} compares when the same item evidence is introduced during inference, testing whether \epic{} improves recommendation by altering the denoising trajectory and preserving target hypotheses rather than merely refining final scores. This matched intervention holds the trained scorer, checkpoint, decoder, beam width, and candidate budget fixed. Post-hoc inference applies the scorer only after token denoising has completed, whereas inline inference feeds the same evidence into every active denoising step. Inline inference improves NDCG@5 over its matched post-hoc counterpart by 15.7\%, 26.6\%, and 26.8\% on \textbf{Beauty}, \textbf{Sports}, and \textbf{Toys}, respectively.

This gap cannot be closed by retuning the post-hoc scoring strength. Control and post-hoc beam signatures remain identical at every audited query-step state, and runtime audits confirm that no token-feedback calls occur in the post-hoc arms. The comparison therefore isolates whether item evidence changes the denoising trajectory, rather than whether a stronger final scorer is used.

The timing sweep further shows that the gain accumulates throughout denoising. Delaying feedback until step 2 reduces NDCG@5 by 4.3--9.3\%, while restricting item evidence to conditional completion leads to an 8.6--14.9\% reduction. Nevertheless, every intermediate timing remains better than pure post-hoc scoring, ruling out a single final scoring pass as the sole source of improvement.

\boldparagraph{Trajectory preservation and hard rescues.}
\label{sec:abl-trajectory-rescue}

We examine how inline feedback changes the search trajectory by following the ground-truth target through each denoising state. A target counts toward trajectory Recall@128 if it remains compatible with at least one active beam and ranks among the top 128 items under the combined beam and conditional completion score. At the final state, inline inference retains more targets on every dataset, yielding net gains of 315/371/360 trajectory hits on
\textbf{Beauty}/\textbf{Sports}/\textbf{Toys}.

Figure~\ref{fig:epic-hard-rescue} decomposes the corresponding changes in top-5 accuracy. A \textit{recovered hit} is a user for whom the control misses the target but inline inference ranks it in the top 5; a \textit{lost hit} is the reverse. Inline inference recovers 261/337/366 hits while losing only 107/106/113, for net gains of 154/231/253 on \textbf{Beauty}/\textbf{Sports}/\textbf{Toys}. In 25/48/49 recovered cases, the target is absent from every final control beam and from the control top-128 pool. These \textit{outside-pool recoveries} cannot be produced by a reranker restricted to the final control candidates. For the remaining recovered hits, the target is still accessible to the control, and the gain may reflect either better trajectory preservation or better ranking.

The outside-pool recoveries show that inline feedback acts before final ranking. In these cases, the target is absent from the final control candidate pool and therefore cannot be recovered by post-hoc reranking. Inline feedback instead changes the denoising trajectory, preserving the target as successive SID positions narrow the feasible candidate set.

\subsection{What Drives the Improvement?} \label{sec:ablation}
\noindent\boldparagraph{Capability ladder and component removals.}
\label{sec:abl-capability}
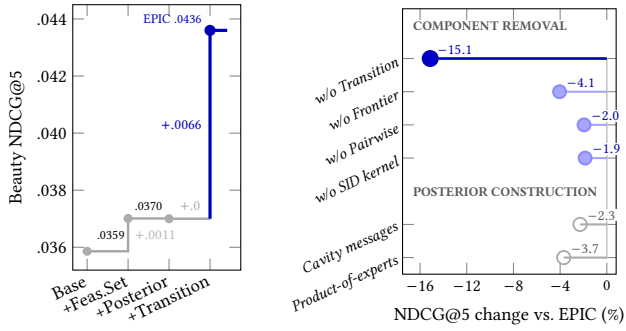
\begin{figure}[t]
\centering
\begin{subfigure}[b]{0.44\linewidth}
  \scriptsize
  \centering
  \begin{tikzpicture}
    \begin{axis}[
      width=1.02\linewidth,
      height=0.70*\axisdefaultheight,
      ymin=.0352,ymax=.0445,
      ylabel={Beauty NDCG@5},
      xmin=-.35,xmax=3.75,
      xtick={0,1,2,3},
      xticklabels={Base,+Feas.Set,+Posterior,+Transition},
      xticklabel style={font=\footnotesize,rotate=25,anchor=east,yshift=-3pt},
      ytick={.036,.038,.040,.042,.044},
      yticklabels={.036,.038,.040,.042,.044},
      scaled y ticks=false,
      yticklabel style={font=\footnotesize},
      tick label style={font=\footnotesize},
      label style={font=\footnotesize},
      clip=false,
    ]
      \draw[gray!70,line width=.9pt]
        (axis cs:0,.03586) -- (axis cs:1,.03586) --
        (axis cs:1,.03701) -- (axis cs:2,.03701) --
        (axis cs:2,.03700) -- (axis cs:3,.03700);
      \draw[blue!72!black,line width=1.2pt]
        (axis cs:3,.03700) -- (axis cs:3,.04360) --
        (axis cs:3.42,.04360);

      \fill[gray!65] (axis cs:0,.03586) circle (1.7pt);
      \fill[gray!65] (axis cs:1,.03701) circle (1.7pt);
      \fill[gray!65] (axis cs:2,.03700) circle (1.7pt);
      \fill[blue!72!black] (axis cs:3,.04360) circle (2.1pt);

      \node[font=\tiny,anchor=south west] at (axis cs:.10,.03608) {.0359};
      \node[font=\tiny,anchor=south] at (axis cs:1.50,.03708) {.0370};
      \node[font=\tiny,text=gray!55,anchor=west]
        at (axis cs:1.03,.03643) {$+.0011$};
      \node[font=\tiny,text=gray!65,anchor=south] at (axis cs:2.52,.03708) {$+.0$};
      \node[font=\tiny,text=blue!72!black,anchor=east]
        at (axis cs:2.94,.04030) {$+.0066$};
      \node[font=\tiny,anchor=south east,text=blue!72!black]
        at (axis cs:2.94,.04366) {EPIC .0436};
    \end{axis}
  \end{tikzpicture}
  \caption{Personalized transitions drive item-to-token feedback.}
  \label{fig:epic-capability-ladder}
\end{subfigure}%
\hspace{0.03\linewidth}%
\begin{subfigure}[b]{0.52\linewidth}
  \scriptsize
  \centering
  \begin{tikzpicture}
    \tikzset{
      design value/.style={font=\tiny,fill=white,inner sep=.5pt,
                           anchor=south west,xshift=2pt,yshift=1pt},
      component value/.style={design value,text=blue!60!black},
      transition value/.style={design value,text=blue!72!black},
      posterior value/.style={design value,text=gray!70!black},
    }
    \begin{axis}[
      width=\linewidth,
      height=0.70*\axisdefaultheight,
      xmin=-17.5,xmax=.8,
      xtick={-16,-12,-8,-4,0},
      xticklabels={$-16$,$-12$,$-8$,$-4$,0},
      xticklabel style={font=\scriptsize},
      xlabel={NDCG@5 change vs. EPIC (\%)},
      xlabel style={font=\footnotesize,yshift=-2pt},
      ytick={1,2,3,4,6,7},
      yticklabels={w/o Transition,w/o Frontier,w/o Pairwise,w/o SID kernel,
                   Cavity messages,Product-of-experts},
      ymin=-.5,ymax=7.5,
      y dir=reverse,
      tick label style={font=\footnotesize},
      yticklabel style={font=\scriptsize\itshape,rotate=25,anchor=east},
      label style={font=\footnotesize},
      clip=false,
    ]
      \draw[black!22,line width=.6pt]
        (axis cs:0,1) -- (axis cs:0,7);

      \addplot+[xcomb,blue!72!black,line width=1pt,mark=*,mark size=2.7pt]
        coordinates {(-15.13,1)};
      \addplot+[xcomb,blue!35,line width=.8pt,mark=*,mark size=2.4pt,
        mark options={fill=blue!35,draw=blue!45}]
        coordinates {(-4.05,2)
                     (-1.95,3)
                     (-1.86,4)};

      \addplot+[xcomb,gray!65,line width=.8pt,mark=o,mark size=2.5pt,
        mark options={fill=white,draw=gray!65,line width=.7pt}]
        coordinates {(-2.29,6)
                     (-3.69,7)};

      \node[font=\tiny\bfseries,text=gray!70!black,anchor=west]
        at (axis cs:-17.1,0) {COMPONENT REMOVAL};
      \node[font=\tiny\bfseries,text=gray!70!black,anchor=west]
        at (axis cs:-17.1,5) {POSTERIOR CONSTRUCTION};

      \node[transition value] at (axis cs:-15.13,1) {$-15.1$};
      \node[component value] at (axis cs:-4.05,2) {$-4.1$};
      \node[component value] at (axis cs:-1.95,3) {$-2.0$};
      \node[component value] at (axis cs:-1.86,4) {$-1.9$};
      \node[posterior value] at (axis cs:-2.29,6) {$-2.3$};
      \node[posterior value] at (axis cs:-3.69,7) {$-3.7$};
    \end{axis}
  \end{tikzpicture}
  \caption{Item-posterior design ablations on Beauty.}
  \label{fig:epic-component-drop}
\end{subfigure}
\caption{The capability progression and matched design ablations locate the gain: personalized transition dominates, while separated item scoring outperforms token-coupled posterior constructions.}
\label{fig:epic-gain-sources}
\vspace{-8pt}
\end{figure}

We next separate gains due to catalog-valid decoding from gains due to explicit item reasoning. Figure~\ref{fig:epic-capability-ladder} constructs this capability ladder under the same \textbf{Beauty} split and evaluation protocol. Feasible-candidate-set fusion modestly improves the unmodified \texttt{LLaDA-Rec} decoder, but adding a generic item posterior without transition memory is neutral relative to this decode-matched control. The clear improvement appears only when the posterior is conditioned on the user's transition history. This indicates that adding a catalog-level scoring head alone does not explain the observed gain.

The component-removal group in Figure~\ref{fig:epic-component-drop} leads to the same conclusion. Removing transition memory causes a 15.13\% NDCG@5 drop and returns the model to the decode-control regime. Frontier supervision is the second-largest isolated component contribution (4.05\%), while pairwise item structure and the SID kernel provide smaller but consistent gains.



\noindent\boldparagraph{Is the transition signal personalized?}
\label{sec:abl-personalization}

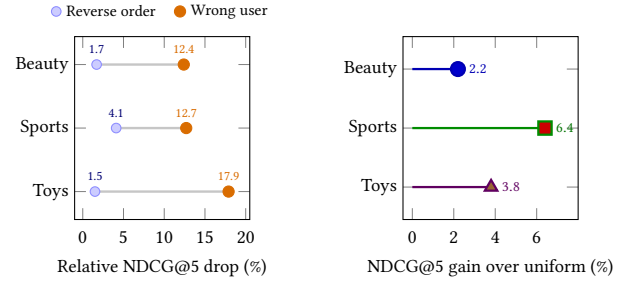
\begin{figure}[t]
\centering
\begin{subfigure}[b]{0.46\linewidth}
  \scriptsize
  \centering
  \begin{tikzpicture}
    \begin{axis}[
      width=\linewidth,
      height=0.55*\axisdefaultheight,
      xmin=-1,xmax=20.5,
      xtick={0,5,10,15,20},
      xticklabel style={font=\footnotesize,yshift=-2pt},
      xlabel={Relative NDCG@5 drop (\%)},
      xlabel style={font=\footnotesize,yshift=-2pt},
      symbolic y coords={Beauty,Sports,Toys},
      ytick={Beauty,Sports,Toys},
      y dir=reverse,
      tick label style={font=\footnotesize},
      label style={font=\footnotesize},
      enlarge y limits=.22,
      legend style={at={(0.5,1.04)},anchor=south,legend columns=2,
                    font=\scriptsize,draw=none,fill=none,
                    /tikz/every even column/.append style={column sep=5pt}},
      clip=false,
    ]
      \addplot+[gray!45,line width=.9pt,no marks,forget plot]
        coordinates {(1.7,Beauty)(12.4,Beauty)};
      \addplot+[gray!45,line width=.9pt,no marks,forget plot]
        coordinates {(4.1,Sports)(12.7,Sports)};
      \addplot+[gray!45,line width=.9pt,no marks,forget plot]
        coordinates {(1.5,Toys)(17.9,Toys)};

      \addplot+[only marks,mark=*,mark size=1.8pt,
        mark options={fill=blue!20,draw=blue!45},
        nodes near coords,point meta=explicit symbolic,
        nodes near coords style={font=\tiny,text=blue!45!black,
                                 anchor=south,yshift=2pt}]
        coordinates {(1.7,Beauty)[1.7](4.1,Sports)[4.1](1.5,Toys)[1.5]};
      \addlegendentry{Reverse order}
      \addplot+[only marks,mark=*,mark size=2.1pt,
        mark options={fill=orange!85!black,draw=orange!85!black},
        nodes near coords,point meta=explicit symbolic,
        nodes near coords style={font=\tiny,text=orange!85!black,
                                 anchor=south,yshift=2pt}]
        coordinates {(12.4,Beauty)[12.4](12.7,Sports)[12.7]
                     (17.9,Toys)[17.9]};
      \addlegendentry{Wrong user}
    \end{axis}
  \end{tikzpicture}
  \caption{User identity matters more than temporal order.}
  \label{fig:epic-history-intervention}
\end{subfigure}%
\hspace{0.06\linewidth}%
\begin{subfigure}[b]{0.46\linewidth}
  \scriptsize
  \centering
  \begin{tikzpicture}
    \begin{axis}[
      width=\linewidth,
      height=0.55*\axisdefaultheight,
      xmin=-.45,xmax=8,
      xtick={0,2,4,6},
      xticklabel style={font=\footnotesize,yshift=-2pt},
      xlabel={NDCG@5 gain over uniform (\%)},
      xlabel style={font=\footnotesize,yshift=-2pt},
      ytick={1,2,3},
      yticklabels={Beauty,Sports,Toys},
      ymin=.45,ymax=3.55,
      y dir=reverse,
      tick label style={font=\footnotesize},
      label style={font=\footnotesize},
      clip=false,
    ]
      \addplot+[xcomb,blue!70!black,line width=.85pt,
        mark=*,mark size=2.6pt]
        coordinates {(2.2,1)};
      \addplot+[xcomb,green!55!black,line width=.85pt,
        mark=square*,mark size=2.5pt]
        coordinates {(6.4,2)};
      \addplot+[xcomb,violet!75!black,line width=.85pt,
        mark=triangle*,mark size=2.7pt]
        coordinates {(3.8,3)};
      \node[font=\tiny,text=blue!70!black,anchor=west,xshift=2pt]
        at (axis cs:2.2,1) {$2.2$};
      \node[font=\tiny,text=green!45!black,anchor=west,xshift=2pt]
        at (axis cs:6.4,2) {$6.4$};
      \node[font=\tiny,text=violet!75!black,anchor=west,xshift=2pt]
        at (axis cs:3.8,3) {$3.8$};
    \end{axis}
  \end{tikzpicture}
  \caption{Learned weighting improves all three datasets.}
  \label{fig:epic-history-weighting}
\end{subfigure}
\caption{EPIC depends on user-specific history; learned position weights
further improve history aggregation.}
\label{fig:epic-personalization}
\vspace{-8pt}
\end{figure}

We determine whether transition memory captures user-specific evidence or merely a dataset-level popularity prior. Replacing a user's history with another user's history reduces NDCG@5 by 12.4--17.9\% across the three datasets (Figure~\ref{fig:epic-history-intervention}). Reversing the correct history preserves its item set but destroys temporal order and yields a smaller, consistent 1.5--4.1\% degradation. Thus, user identity explains most of the transition gain, while temporal order provides additional signal.

The transition module also benefits from learning which history positions to trust. At each dataset's default history window, learned weighting improves over uniform aggregation by 2.2--6.4\% (Figure~\ref{fig:epic-history-weighting}), with the largest gain on \textbf{Sports}. This intervention establishes that position-dependent weighting is useful.

\noindent\boldparagraph{Posterior construction.}
\label{sec:abl-posterior-construction}

We examine whether the item posterior should be learned independently of the backbone's token evidence or derived from it. In the principal model, token likelihood forms the candidate support and contributes at final fusion, but is not added to the posterior energy (Section~\ref{sec:explicit-item-posterior}). Two alternative constructions fold token evidence directly into the exponent: a product-of-experts prior that adds the token-induced tuple score of Equation~\eqref{eq:token-tuple-score} to $E_t(i)$, and leave-one-position-out cavity messages (Figure~\ref{fig:epic-component-drop}). Both are consistently worse than the independently learned scorer on \textbf{Beauty}, by 3.7\% and 2.3\% validation NDCG@5, respectively.

These results support separating candidate-support construction from item-preference estimation. Backbone likelihood provides a tractable support for item inference and remains available at final fusion, whereas including it again in the item-posterior energy degrades validation ranking. This finding is consistent with prior evidence that sequence likelihood need not be calibrated for item elevance~\cite{tikhonovich2026gryphon, zhao2026simgr}. Accordingly, Equation~\eqref{eq:candidate-support} treats preselection as a non-differentiable support operation, and the posterior energy is learned without an additional token-score term. This prevents token likelihood from being reused as a second preference signal within the item posterior.

\subsection{When Is EPIC Effective, and at What Cost?} \label{sec:effective}
\noindent\boldparagraph{Partial SID resolution and ambiguity.}
\label{sec:abl-denoising-sufficiency}

\begin{figure}[t]
\centering
\begin{subfigure}[b]{0.46\linewidth}
  \scriptsize
  \centering
  \vspace*{-10pt}
  \begin{tikzpicture}
    \tikzset{
      sid label base/.style={font=\tiny,text=gray!75!black},
      sid label 0/.style={sid label base,anchor=south west,xshift=2pt,yshift=2pt},
      sid label 1/.style={sid label base,anchor=south east,xshift=-2pt,yshift=2pt},
      sid label 2/.style={font=\tiny,text=blue!70!black,anchor=north east,
                         xshift=-2pt,yshift=-2pt},
      sid label 3/.style={sid label base,anchor=south,xshift=0pt,yshift=2pt},
      sid label 4/.style={sid label base,anchor=south east,xshift=-2pt,yshift=2pt},
    }
    \begin{axis}[
      width=\linewidth,
      height=0.55*\axisdefaultheight,
      xmin=25,xmax=155,
      ymin=.0335,ymax=.045,
      xlabel={Latency (ms/user)},
      ylabel={Beauty test NDCG@5},
      tick label style={font=\footnotesize},
      label style={font=\footnotesize},
      scaled y ticks=false,
      ytick={.035,.040,.045},
      yticklabel style={/pgf/number format/fixed,
                        /pgf/number format/precision=3,font=\footnotesize},
      clip=false,
    ]
      \addplot+[gray!65,line width=.7pt,no marks]
        coordinates {(31.30,.03481)(62.36,.04204)(85.82,.04385)
                     (114.51,.04360)(147.85,.04358)};
      \addplot+[only marks,gray!75,mark=o,mark size=2.7pt,
                mark options={fill=white,line width=.7pt}]
        coordinates {(31.30,.03481)(62.36,.04204)
                     (114.51,.04360)(147.85,.04358)};
      \addplot+[only marks,blue!70!black,mark=*,mark size=3.2pt]
        coordinates {(85.82,.04385)};

      \node[sid label 0] at (axis cs:31.30,.03481) {0 SID};
      \node[sid label 1] at (axis cs:68.36,.04204) {1 SID};
      \node[sid label 2] at (axis cs:105.82,.04385)
        {\shortstack{\textbf{2 SID}\\[-1pt]best}};
      \node[sid label 3] at (axis cs:114.51,.04160) {3 SID};
      \node[sid label 4] at (axis cs:162.85,.04158) {4 SID};
    \end{axis}
  \end{tikzpicture}
  \par\vspace*{10pt}
  \caption{Two resolved SID positions give the best quality--latency operating point.}
  \label{fig:epic-sid-pareto}
\end{subfigure}%
\hspace{0.06\linewidth}%
\begin{subfigure}[b]{0.46\linewidth}
  \scriptsize
  \centering
  \begin{tikzpicture}
    \begin{axis}[
      width=\linewidth,
      height=0.55*\axisdefaultheight,
      ylabel={EPIC $-$ control NDCG@5},
      symbolic x coords={1,2--4,5--16,17--64},
      xtick={1,2--4,5--16,17--64},
      xlabel={Final $|\mathcal C|$ bin},
      xticklabel style={font=\footnotesize,rotate=25,anchor=north east,
                        yshift=-2pt},
      xlabel style={font=\footnotesize,yshift=0pt},
      tick label style={font=\footnotesize},
      label style={font=\footnotesize},
      ymin=0,ymax=.034,
      scaled y ticks=false,
      ytick={0,.01,.02,.03},
      yticklabels={0,.01,.02,.03},
      yticklabel style={font=\footnotesize},
      xmajorgrids=false,
      legend style={at={(0.5,1.05)},anchor=south,legend columns=3,
                    font=\scriptsize,draw=none,fill=none},
      every axis plot/.append style={line width=.65pt,mark size=2.8pt},
    ]
      \addplot+[blue!70!black,mark=*] coordinates {(1,.00616)(2--4,.00660)(5--16,.01056)(17--64,.00208)};
      \addlegendentry{Beauty}
      \addplot+[green!55!black,mark=square*] coordinates {(1,.00570)(2--4,.00564)(5--16,.00470)(17--64,.00013)};
      \addlegendentry{Sports}
      \addplot+[violet!75!black,mark=triangle*] coordinates {(1,.00919)(2--4,.01299)(5--16,.01236)(17--64,.03122)};
      \addlegendentry{Toys}
      \node[font=\tiny,anchor=south east] at (axis cs:17--64,.03122) {$n=34$};
    \end{axis}
  \end{tikzpicture}
  \caption{Gain is positive across feasible-set strata; the final Toys point is descriptive only.}
  \label{fig:epic-compatible-strata}
\end{subfigure}
\caption{EPIC helps across ambiguity regimes, while partial SID resolution
provides a favorable quality--latency operating point.}
\label{fig:epic-when-and-cost}
\vspace{-10pt}
\end{figure}
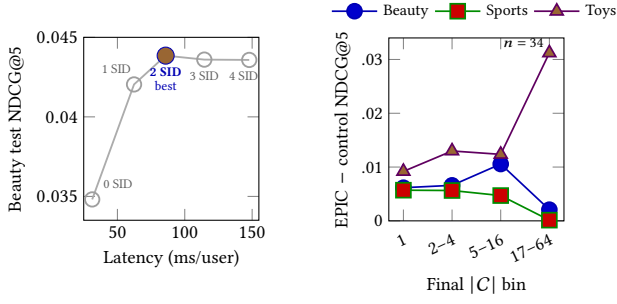

Figure~\ref{fig:epic-sid-pareto} varies the number of SID positions explicitly resolved before catalog ranking. Resolving two positions achieves the best \textbf{Beauty} NDCG@5 (.04385) and is 1.72$\times$ faster than resolving all four. The nearly flat quality from two to four positions suggests that a short partial SID is often sufficient to identify the useful item-hypothesis region; it does not imply that every item is uniquely identified after two positions.

We also stratify gains by the size of the final feasible candidate set on the control trajectory (Figure~\ref{fig:epic-compatible-strata}). \epic{} improves every populated stratum, but ambiguity alone does not guarantee a large gain. \textbf{Sports} receives almost no benefit in its broadest feasible-set bin, while the largest Toys point contains only 34 users and is therefore descriptive. Positive gain at $|\mathcal C|=1$ reflects benefit accumulated at earlier states, before the final set became unique.

\noindent\boldparagraph{Parameter and compute cost.}
\label{sec:abl-efficiency}

\epic{} adds item-level reasoning through parameter-efficient adaptation. On the same Tesla T4, its adapter stage trains only 0.391M parameters, 94.3\% fewer than full \texttt{LLaDA-Rec} fine-tuning, while improving adapter-training throughput by 42.7\% and reducing peak training memory by 55.9\%. This comparison excludes the cost of pretraining the frozen backbone, which is shared by both inference systems. Candidate reasoning increases inference latency by 34.2\%, from 93.0 to 124.8 ms/user, but leaves peak inference memory nearly unchanged. The resolved-SID result in Figure~\ref{fig:epic-sid-pareto} further shows that this latency cost is controllable: resolving two SID positions attains the best observed quality while requiring less latency than the complete trajectory.

\begin{table}[t]
  \caption{Parameter and compute cost on one Tesla T4. Relative changes are
  measured against LLaDA-Rec.}
  \label{tab:abl-efficiency}
  \centering
  \small
  \setlength{\tabcolsep}{3.5pt}
  \begin{tabular}{lrrr}
    \toprule
    Measurement & \texttt{LLaDA-Rec} & \epic{} & Change \\
    \midrule
    \multicolumn{4}{l}{\textit{\underline{Training adaptation}}} \\
    Trainable parameters (M)   & 6.847  & 0.391  & $-94.3\%$ \\
    Throughput (ex/s)          & 333.8  & 476.3  & $+42.7\%$ \\
    Peak memory (MiB)          & 4299.6 & 1895.4 & $-55.9\%$ \\
    \midrule
    \multicolumn{4}{l}{\textit{\underline{Inference}}} \\
    Latency (ms/user)          & 93.0   & 124.8  & $+34.2\%$ \\
    Peak memory (MiB)          & 3359.8 & 3361.3 & $+0.04\%$ \\
    \bottomrule
  \end{tabular}
\end{table}

\subsection{Case Study: A Hard Rescue} \label{sec:case_study}

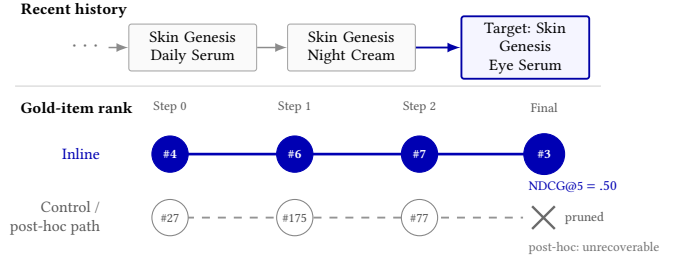
\begin{figure}[t]
\centering
\scriptsize
\begin{tikzpicture}[
  x=1cm,
  y=1cm,
  >=latex,
  item/.style={
    draw=black!35,
    fill=black!2,
    rounded corners=1pt,
    minimum height=.62cm,
    text width=1.55cm,
    align=center,
    inner sep=2pt,
    font=\scriptsize
  },
  target/.style={
    item,
    draw=blue!65!black,
    fill=blue!4,
    line width=.7pt
  },
  inline point/.style={
    circle,
    draw=blue!70!black,
    fill=blue!70!black,
    text=white,
    minimum size=.48cm,
    inner sep=0pt,
    font=\tiny\bfseries
  },
  control point/.style={
    circle,
    draw=black!48,
    fill=white,
    text=black!70,
    minimum size=.48cm,
    inner sep=0pt,
    font=\tiny
  }
]

\node[font=\scriptsize\bfseries,anchor=west] at (0,2.62) {Recent history};
\node[font=\small,text=black!55] (earlier) at (.95,2.13) {$\cdots$};
\node[item] (serum) at (2.35,2.13)
  {Skin Genesis\\Daily Serum};
\node[item] (cream) at (4.45,2.13)
  {Skin Genesis\\Night Cream};
\node[target] (eye) at (6.75,2.13)
  {Target: Skin Genesis\\Eye Serum};
\draw[->,black!45,line width=.55pt] (earlier) -- (serum);
\draw[->,black!45,line width=.55pt] (serum) -- (cream);
\draw[->,blue!65!black,line width=.7pt] (cream) -- (eye);

\draw[black!18] (0,1.62) -- (7.75,1.62);

\node[font=\scriptsize\bfseries,anchor=west] at (0,1.34) {Gold-item rank};

\foreach \x/\lab in {2.05/{Step 0},3.70/{Step 1},5.35/{Step 2},7.00/{Final}} {
  \node[font=\tiny,text=black!65] at (\x,1.34) {\lab};
}

\node[font=\scriptsize,anchor=east,text=blue!70!black]
  at (1.20,.72) {Inline};
\node[font=\scriptsize,anchor=east,text=black!65,align=right]
  at (1.20,-.12) {Control /\\post-hoc path};

\draw[blue!70!black,line width=1pt]
  (2.05,.72) -- (3.70,.72) -- (5.35,.72) -- (7.00,.72);
\node[inline point] at (2.05,.72) {\#4};
\node[inline point] at (3.70,.72) {\#6};
\node[inline point] at (5.35,.72) {\#7};
\node[inline point,minimum size=.54cm] at (7.00,.72) {\#3};

\draw[black!42,dashed,line width=.7pt]
  (2.05,-.12) -- (3.70,-.12) -- (5.35,-.12) -- (6.72,-.12);
\node[control point] at (2.05,-.12) {\#27};
\node[control point] at (3.70,-.12) {\#175};
\node[control point] at (5.35,-.12) {\#77};
\draw[black!58,line width=.9pt]
  (6.84,-.27) -- (7.14,.03)
  (6.84,.03) -- (7.14,-.27);
\node[font=\tiny,anchor=west,text=black!65] at (7.20,-.12) {pruned};

\node[font=\tiny,anchor=west,text=blue!70!black]
  at (6.72,.29) {NDCG@5 $=.50$};
\node[font=\tiny,anchor=west,text=black!55]
  at (6.72,-.52) {post-hoc: unrecoverable};

\end{tikzpicture}
\caption{\textbf{An audited hard rescue on Beauty.} For the first hard rescue in deterministic test order, inline feedback preserves the target through denoising and ranks it third.  The control path prunes the target from every final beam, leaving no candidate for post-hoc scoring to recover.}
\label{fig:epic-quantitative-example}
\vspace{-7pt}
\end{figure}

We conclude the analysis with an audited test instance that illustrates the trajectory-preservation mechanism identified above in Figure~\ref{fig:epic-quantitative-example}. To avoid selecting an example based on semantic appearance, we use the first \textbf{Beauty} hard rescue in deterministic test order. Its two most recent interactions are a Skin Genesis daily serum and night cream, while the target is an eye serum from the same product line. This history makes the relevant transition interpretable without requiring product images or additional metadata.

The quantitative trajectory shows why timing matters. Under control denoising, the target moves from ranks 27 to 175 and 77 before being pruned from every final beam. Pure post-hoc scoring follows this same trajectory and therefore has no surviving target candidate to recover.  With inline item feedback, the target remains available at every state, following ranks 4, 6, and 7 before finishing at rank 3, for an NDCG@5 of .50. This example provides a concrete instance of the elimination mechanism quantified by the aggregate hard-rescue analysis.
\section{Conclusion}

Semantic ID diffusion improves how identifier tokens are generated, but recommendation ultimately requires selecting a complete catalog item. This mismatch creates a token--item inference gap: locally plausible token decisions may prematurely eliminate the preferred item before its item-level evidence is considered. EPIC addresses this gap by explicitly reasoning over the feasible candidates represented by each partial SID. At every denoising step, it compares these candidates using user-specific item evidence and projects the resulting posterior back into token generation, allowing promising items to influence decoding while they remain reachable rather than only reranking the final survivors.

Across four Amazon benchmarks, EPIC improves all 16 dataset--metric combinations, demonstrating consistent benefits across different catalog domains and ranking cutoffs. Matched inline--post-hoc comparisons further show that applying item evidence during denoising is substantially more effective than using the same scorer only after generation. Together with history-perturbation analyses, these results attribute the gains primarily to user-specific transition evidence guiding intermediate decoding decisions, rather than to a generic change in candidate scoring. Future work could replace the current exact catalog scan with approximate candidate construction, adapt the candidate budget across denoising steps, and extend item inference to distinguish catalog items sharing the same SID. 
\section*{Ethical Considerations}
This work uses publicly available, de-identified Amazon review benchmarks and involves neither new data collection nor human-subject intervention. Nevertheless, historical interactions and item metadata may encode popularity, exposure, demographic, and representation biases. If deployed, these biases could be amplified through repeated recommendation and feedback loops, potentially disadvantaging long-tail items or underrepresented user groups. The method should therefore be evaluated beyond ranking accuracy, including subgroup performance, catalog exposure, diversity, and robustness to sparse interaction histories. Interaction logs should be collected with appropriate consent, minimized, securely stored, and protected against re-identification. Sensitive attributes should not be used for personalization without a clear, justified, and lawful purpose. Finally, real-world deployment would require ongoing bias auditing, user controls, and mechanisms for identifying and correcting harmful recommendations.

\newpage

\bibliographystyle{ACM-Reference-Format}
\bibliography{main}

@inproceedings{tang2018caser,
  author    = {Tang, Jiaxi and Wang, Ke},
  title     = {Personalized Top-{N} Sequential Recommendation via Convolutional Sequence Embedding},
  booktitle = {Proceedings of the Eleventh ACM International Conference on Web Search and Data Mining},
  year      = {2018}
}

@inproceedings{ma2019hgn,
  author    = {Ma, Chen and Kang, Peng and Liu, Xue},
  title     = {Hierarchical Gating Networks for Sequential Recommendation},
  booktitle = {Proceedings of the 25th ACM SIGKDD International Conference on Knowledge Discovery and Data Mining},
  year      = {2019}
}

@inproceedings{zhang2019fdsa,
  author    = {Zhang, Tingting and Zhao, Pengpeng and Liu, Yanchi and Sheng, Victor S. and Xu, Jiajie and Wang, Deqing and Liu, Guanfeng and Zhou, Xiaofang},
  title     = {Feature-level Deeper Self-Attention Network for Sequential Recommendation},
  booktitle = {Proceedings of the Twenty-Eighth International Joint Conference on Artificial Intelligence},
  year      = {2019}
}

@inproceedings{kang2018sasrec,
  author    = {Kang, Wang-Cheng and McAuley, Julian},
  title     = {Self-Attentive Sequential Recommendation},
  booktitle = {2018 IEEE International Conference on Data Mining},
  year      = {2018}
}

@inproceedings{sun2019bert4rec,
  author    = {Sun, Fei and Liu, Jun and Wu, Jian and Pei, Changhua and Lin, Xiao and Ou, Wenwu and Jiang, Peng},
  title     = {{BERT4Rec}: Sequential Recommendation with Bidirectional Encoder Representations from Transformer},
  booktitle = {Proceedings of the 28th ACM International Conference on Information and Knowledge Management},
  year      = {2019}
}

@inproceedings{hidasi2016gru4rec,
  author    = {Hidasi, Bal{\'a}zs and Karatzoglou, Alexandros and Baltrunas, Linas and Tikk, Domonkos},
  title     = {Session-based Recommendations with Recurrent Neural Networks},
  booktitle = {Proceedings of the International Conference on Learning Representations},
  year      = {2016}
}

@inproceedings{zhou2020s3rec,
  author    = {Zhou, Kun and Wang, Hui and Zhao, Wayne Xin and Zhu, Yutao and Wang, Sirui and Zhang, Fuzheng and Wang, Zhongyuan and Wen, Ji-Rong},
  title     = {{S3-Rec}: Self-Supervised Learning for Sequential Recommendation with Mutual Information Maximization},
  booktitle = {Proceedings of the 29th ACM International Conference on Information and Knowledge Management},
  year      = {2020}
}

@inproceedings{cao2021genre,
  author    = {De Cao, Nicola and Izacard, Gautier and Riedel, Sebastian and Petroni, Fabio},
  title     = {Autoregressive Entity Retrieval},
  booktitle = {Proceedings of the International Conference on Learning Representations},
  year      = {2021}
}

@inproceedings{geng2022recommendation,
  author    = {Geng, Shijie and Liu, Shuchang and Fu, Zuohui and Ge, Yingqiang and Zhang, Yongfeng},
  title     = {Recommendation as Language Processing ({RLP}): A Unified Pretrain, Personalized Prompt \& Predict Paradigm ({P5})},
  booktitle = {Proceedings of the 16th ACM Conference on Recommender Systems},
  year      = {2022}
}

@inproceedings{tay2022dsi,
  author    = {Tay, Yi and Tran, Vinh Q. and Dehghani, Mostafa and Ni, Jianmo and Bahri, Dara and Mehta, Harsh and Qin, Zhen and Hui, Kai and Zhao, Zhe and Gupta, Jai and Schuster, Tal and Cohen, William W. and Metzler, Donald},
  title     = {Transformer Memory as a Differentiable Search Index},
  booktitle = {Advances in Neural Information Processing Systems},
  year      = {2022}
}

@inproceedings{rajput2023tiger,
  author    = {Rajput, Shashank and Mehta, Nikhil and Singh, Anima and Keshavan, Raghunandan H. and Vu, Trung and Heldt, Lukasz and Hong, Lichan and Tay, Yi and Tran, Vinh Q. and Samost, Jonah and Kula, Maciej and Chi, Ed H. and Sathiamoorthy, Maheswaran},
  title     = {Recommender Systems with Generative Retrieval},
  booktitle = {Advances in Neural Information Processing Systems},
  year      = {2023}
}

@inproceedings{hou2023vqrec,
  author    = {Hou, Yupeng and He, Zhankui and McAuley, Julian and Zhao, Wayne Xin},
  title     = {Learning Vector-Quantized Item Representation for Transferable Sequential Recommenders},
  booktitle = {Proceedings of the ACM Web Conference 2023},
  year      = {2023}
}

@inproceedings{wang2024letter,
  author    = {Wang, Wenjie and Bao, Honghui and Lin, Xinyu and Zhang, Jizhi and Li, Yongqi and Feng, Fuli and Ng, See-Kiong and Chua, Tat-Seng},
  title     = {Learnable Item Tokenization for Generative Recommendation},
  booktitle = {Proceedings of the 33rd ACM International Conference on Information and Knowledge Management},
  year      = {2024}
}

@inproceedings{zheng2024lcrec,
  author    = {Zheng, Bowen and Hou, Yupeng and Lu, Hongyu and Chen, Yu and Zhao, Wayne Xin and Chen, Ming and Wen, Ji-Rong},
  title     = {Adapting Large Language Models by Integrating Collaborative Semantics for Recommendation},
  booktitle = {2024 IEEE 40th International Conference on Data Engineering},
  year      = {2024}
}

@inproceedings{zhai2024hstu,
  author    = {Zhai, Jiaqi and Liao, Lucy and Liu, Xing and Wang, Yueming and Li, Rui and Cao, Xuan and Gao, Leon and Gong, Zhaojie and Gu, Fangda and He, Jiayuan and Lu, Yinghai and Shi, Yu},
  title     = {Actions Speak Louder than Words: Trillion-Parameter Sequential Transducers for Generative Recommendations},
  booktitle = {Proceedings of the 41st International Conference on Machine Learning},
  year      = {2024}
}

@inproceedings{hou2025actionpiece,
  author    = {Hou, Yupeng and Ni, Jianmo and He, Zhankui and Sachdeva, Noveen and Kang, Wang-Cheng and Chi, Ed H. and McAuley, Julian and Cheng, Derek Zhiyuan},
  title     = {{ActionPiece}: Contextually Tokenizing Action Sequences for Generative Recommendation},
  booktitle = {Proceedings of the 42nd International Conference on Machine Learning},
  year      = {2025}
}

@inproceedings{hou2025rpg,
  author    = {Hou, Yupeng and Li, Jiacheng and Shin, Ashley and Jeon, Jinsung and Santhanam, Abhishek and Shao, Wei and Hassani, Kaveh and Yao, Ning and McAuley, Julian},
  title     = {Generating Long Semantic {IDs} in Parallel for Recommendation},
  booktitle = {Proceedings of the 31st ACM SIGKDD Conference on Knowledge Discovery and Data Mining V.2},
  year      = {2025}
}

@inproceedings{oord2017vqvae,
  author    = {van den Oord, A{\"a}ron and Vinyals, Oriol and Kavukcuoglu, Koray},
  title     = {Neural Discrete Representation Learning},
  booktitle = {Advances in Neural Information Processing Systems},
  year      = {2017}
}

@inproceedings{ni2022sentencet5,
  author    = {Ni, Jianmo and Hernandez Abrego, Gustavo and Constant, Noah and Ma, Ji and Hall, Keith B. and Cer, Daniel and Yang, Yinfei},
  title     = {{Sentence-T5}: Scalable Sentence Encoders from Pre-trained Text-to-Text Models},
  booktitle = {Findings of the Association for Computational Linguistics: ACL 2022},
  year      = {2022}
}

@inproceedings{hou2026bridging,
  author    = {Hou, Yupeng and Li, Jiacheng and Fu, Xiangjun and He, Zhankui and Yan, An and Chen, Xiusi and McAuley, Julian},
  title     = {Bridging Language and Items for Retrieval and Recommendation: Benchmarking {LLM}s as Semantic Encoders},
  booktitle = {Proceedings of the 64th Annual Meeting of the Association for Computational Linguistics (Volume 1: Long Papers)},
  year      = {2026}
}

@inproceedings{austin2021d3pm,
  author    = {Austin, Jacob and Johnson, Daniel D. and Ho, Jonathan and Tarlow, Daniel and van den Berg, Rianne},
  title     = {Structured Denoising Diffusion Models in Discrete State-Spaces},
  booktitle = {Advances in Neural Information Processing Systems},
  year      = {2021}
}

@inproceedings{nie2025llada,
  author    = {Nie, Shen and Zhu, Fengqi and You, Zebin and Zhang, Xiaolu and Ou, Jingyang and Hu, Jun and Zhou, Jun and Lin, Yankai and Wen, Ji-Rong and Li, Chongxuan},
  title     = {Large Language Diffusion Models},
  booktitle = {Advances in Neural Information Processing Systems},
  year      = {2025}
}

@inproceedings{wang2023diffrec,
  author    = {Wang, Wenjie and Xu, Yiyan and Feng, Fuli and Lin, Xinyu and He, Xiangnan and Chua, Tat-Seng},
  title     = {Diffusion Recommender Model},
  booktitle = {Proceedings of the 46th International ACM SIGIR Conference on Research and Development in Information Retrieval},
  year      = {2023}
}

@inproceedings{yang2023dreamrec,
  author    = {Yang, Zhengyi and Wu, Jiancan and Wang, Zhicai and Wang, Xiang and Yuan, Yancheng and He, Xiangnan},
  title     = {Generate What You Prefer: Reshaping Sequential Recommendation via Guided Diffusion},
  booktitle = {Advances in Neural Information Processing Systems},
  year      = {2023}
}

@article{shi2025lladarec,
  author        = {Shi, Teng and Shen, Chenglei and Yu, Weijie and Nie, Shen and Li, Chongxuan and Zhang, Xiao and He, Ming and Han, Yan and Xu, Jun},
  title         = {{LLaDA-Rec}: Discrete Diffusion for Parallel Semantic {ID} Generation in Generative Recommendation},
  journal       = {arXiv preprint arXiv:2511.06254},
  year          = {2025}
}

@inproceedings{liu2026diffgrm,
  author    = {Liu, Zhao and Zhu, Yichen and Yang, Yiqing and Lv, Xiao and Tang, Guoping and Huang, Rui and Luo, Qiang and Tang, Ruiming and Gai, Kun and Zhou, Guorui},
  title     = {{DiffGRM}: Diffusion-based Generative Recommendation Model},
  booktitle = {Proceedings of the ACM Web Conference 2026},
  year      = {2026}
}

@article{shah2025maskgr,
  author        = {Shah, Kulin and Kumar, Bhuvesh and Shah, Neil and Collins, Liam},
  title         = {Masked Diffusion for Generative Recommendation},
  journal       = {arXiv preprint arXiv:2511.23021},
  year          = {2025}
}

@article{mu2026mdgr,
  author        = {Mu, Lingyu and Deng, Hao and Xing, Haibo and Hu, Jinxin and Zhang, Yu and Zeng, Xiaoyi and Zhang, Jing},
  title         = {Masked Diffusion Generative Recommendation},
  journal       = {arXiv preprint arXiv:2601.19501},
  year          = {2026}
}

@inproceedings{wei2026mhl,
  author    = {Wei, Kaiwen and He, Kejun and Kang, Xiaomian and Zhang, Jie and Yang, Yuming and Jin, Li and Li, Zhenyang and Zhong, Jiang and Bai, He and Zhu, Junnan},
  title     = {From Past To Path: Masked History Learning for Next-Item Prediction in Generative Recommendation},
  booktitle = {Proceedings of the 64th Annual Meeting of the Association for Computational Linguistics (Volume 1: Long Papers)},
  year      = {2026}
}

@inproceedings{zheng2025utgrec,
  author    = {Zheng, Bowen and Lu, Hongyu and Chen, Yu and Zhao, Wayne Xin and Wen, Ji-Rong},
  title     = {Universal Item Tokenization for Transferable Generative Recommendation},
  booktitle = {Proceedings of the 49th International ACM SIGIR Conference on Research and Development in Information Retrieval},
  year      = {2026}
}

@article{zhong2025pctx,
  author        = {Zhong, Qiyong and Su, Jiajie and Ma, Yunshan and McAuley, Julian and Hou, Yupeng},
  title         = {{Pctx}: Tokenizing Personalized Context for Generative Recommendation},
  journal       = {arXiv preprint arXiv:2510.21276},
  year          = {2025}
}

@article{zhao2026simgr,
  author        = {Zhao, Yuanbo and Liu, Ruochen and Wang, Senzhang and Yin, Jun and Dong, Yuxin and Gong, Huan and Chen, Hao and Pan, Shirui and Zhang, Chengqi},
  title         = {{SimGR}: Escaping the Pitfalls of Generative Decoding in {LLM}-based Recommendation},
  journal       = {arXiv preprint arXiv:2602.07847},
  year          = {2026}
}

@article{tikhonovich2026gryphon,
  author        = {Tikhonovich, Daria and Sorokin, Oleg and Dodonov, Vladislav and Ulianova, Mariia and Murzin, Ilya},
  title         = {{Gryphon}: A Unified Architecture for Semantic-{ID} Generation and Item-Level Scoring in Industrial Recommendations},
  journal       = {arXiv preprint arXiv:2606.08604},
  year          = {2026}
}

@article{wang2026understanding,
  author        = {Wang, Junting and He, Xinrui and Li, Yunzhe and Sundaram, Hari},
  title         = {Understanding Semantic {IDs}: From Item Representation to Item Selection in Generative Recommendation},
  journal       = {arXiv preprint arXiv:2607.24995},
  year          = {2026}
}



\end{document}